\documentclass[twocolumn]{aastex61}
\pdfoutput=1 
\usepackage{amsmath,amstext}
\usepackage[T1]{fontenc}
\usepackage{apjfonts} 
\usepackage[figure,figure*]{hypcap}
\usepackage{hyperref}  
\usepackage{booktabs}
\usepackage{multirow}
\usepackage{graphicx}

\newcommand{\nver}{\hat{\mathbf{n}}} 

\shorttitle{Broadband SEDs of SDSS-selected Quasars and of their host galaxies}
\shortauthors{F. Bianchini et al.}

\begin{document}

\title{Broadband spectral energy distributions of SDSS-selected Quasars and of their host galaxies: intense activity at the onset of AGN feedback}

\author{Federico Bianchini}
\affiliation{School of Physics, University of Melbourne, Parkville, VIC 3010, Australia}
\email{fbianchini@unimelb.edu.au}
\author{Giulio Fabbian}
\affiliation{Institut d'Astrophysique Spatiale, CNRS (UMR 8617), Univ. Paris-Sud, Universit\'{e} Paris-Saclay, B\^{a}t. 121, 91405 Orsay, France}
\author{Andrea Lapi}
\affiliation{SISSA, Via Bonomea 265, 34136 Trieste, Italy}
\affiliation{INFN-Sezione di Trieste, via Valerio 2, I-34127 Trieste, Italy}
\affiliation{INAF-Osservatorio Astronomico di Trieste, via Tiepolo 11, I-34131 Trieste, Italy}
\author{Joaquin Gonzalez-Nuevo}
\affiliation{Departamento de F\`isica, Universidad de Oviedo, C. Federico Garc\`ia Lorca 18, 33007 Oviedo, Spain}
\author{Roberto Gilli}
\affiliation{INAF-Osservatorio di Astrofisica e Scienza dello Spazio di Bologna, 
Via Gobetti 93/3 - 40129 Bologna, Italy}
\author{Carlo Baccigalupi}
\affiliation{SISSA, Via Bonomea 265, 34136 Trieste, Italy}
\affiliation{INFN-Sezione di Trieste, via Valerio 2, I-34127 Trieste, Italy}

\begin{abstract}
We present the mean spectral energy distribution (SED) of a sample of optically selected quasars (QSOs) at redshifts of $1 \le z \le 5$. To derive it, we exploit photometric information from SDSS, UKIDSS, and WISE surveys in combination with a stacking analysis of \textit{Herschel}, \textit{AKARI}, and \textit{Planck} maps at the location of the QSOs. The near-UV and optical parts of the reconstructed mean rest-frame SED are similar to those found in other studies. However, the SED shows an excess at 1-2 $\mu$m (when compared to the aforementioned SEDs normalized in the near-UV) and a prominent bump around 4-6 $\mu$m, followed by a decrease out to $\sim 20 \,\mu$m and a subsequent far-IR bump. From the fitted SEDs we estimate the average active galactic nuclei (AGN) luminosity $L_{\rm AGN}$ and star formation rate (SFR) as function of cosmic time, finding typical $L_{\rm AGN} \sim 10^{46} - 10^{47}$ erg/s and SFR $\sim 50 - 1000\, M_{\odot}/$yr.  We develop mid-IR based criteria to split the QSO sample, finding that these allow us to move along the average relationship in the SFR vs. $L_{\rm AGN}$ diagram toward increasing AGN luminosities. When interpreted in the context of the in-situ coevolution scenario presented by \citet{Lapi2014}, our results suggest that the detection in the far-IR band is an effective criterion to select objects where the star formation is on the verge of being affected by energy/momentum feedback from the central AGN.
\end{abstract}


\section{Introduction}

According to our current understanding of galaxy formation and evolution, the nuclei of active galaxies (AGN) are powered by accretion of matter onto supermassive black holes \citep[SMBHs]{Rees1984}, which makes them among the most luminous extragalactic sources \citep{Soltan1982}. The observed correlations between the black hole mass $M_{\rm BH}$ and the properties of the host galaxy \citep[$M_{\rm BH}-\sigma$ relation]{Ferrarese2000,Gebhardt2000}, \citep[$M_{\rm BH}-M_{\rm bulge}$]{Kormendy1995,Magorrian1998} suggest the existence of a coupling between black hole (BH) growth and star formation \citep[e.g.][]{Lapi2014}. A further observation that corroborates this picture is that both the cosmic star formation and the black hole accretion histories peak around $z\sim 2$ \citep{Aird2010,Madau2014}. 

In the standard evolutionary framework the host galaxy and the SMBH grow coevally. The galaxy undergo phases of intense starburst activity while the central SMBH grows by accretion, until the AGN driven large-scale outflows expel the gas and dust from the interstellar medium \citep[e.g.][]{King2003,dimatteo2005}. The depletion of gas regulates the accretion of matter onto the SMBH and may eventually lead to suppression of the star formation activity, leaving a red and dead galaxy behind \citep[e.g.][]{Granato2003,Croton2006}. 

The specific interplay between AGN feedback and the surrounding material plays an important role in the evolution of the host galaxy, as it is thought to inhibit star formation and reheat the interstellar and circumgalactic medium \citep[e.g.][]{Sturm2011,Fabian2012,Cicone2014}. AGN feedback also leaves an imprint on the large-scale structure (LSS) of the universe, as it affects the small-scale clustering of dark matter \citep{vanDaalen2011}, potentially biasing the cosmological analysis of clustering and weak lensing data \citep{Semboloni2011,Eifler2015}. Despite its key role in galaxy evolution, the details of the mechanism and energetics behind AGN feedback are still poorly understood \citep[see][for reviews]{Alexander2012,Fabian2012}.

The AGN spectral energy distribution (SED), which spans the entire electromagnetic spectrum from the radio band to the X-rays, provides a key tool to unravel the nature of the physical processes that govern the SBMH/host-galaxy co-evolution. For example, based on the simple schematic provided by the unified AGN model \citep{Antonucci1993,Urry1995}, the dust produced during the star-forming phase is supposed to form a torus surrounding the accretion disk that reprocesses most of the ultraviolet (UV) and optical photons into the infrared (IR) band. Since the AGN can contribute to the IR luminosity due to the thermal radiation re-emitted by the surrounding torus, it is clear that disentangling the contributions to the total SED from the accretion disk, the dusty torus, and the host-galaxy is of pivotal relevance \cite[e.g.][]{Mullaney2011,delmoro2013,Leipski2013,Delvecchio2014,Netzer2016,Symeonidis2016,Stanley2017}.

The aim of this paper is to provide a multi-wavelength picture of AGN by reconstructing the mean SED of QSOs from the near-UV/optical to the far-IR. To this end, we use the optically selected QSOs from the Sloan Digital Sky Survey (SDSS). By means of template fitting, we derive constraints on the AGN bolometric luminosity and star formation rate (SFR) in the host galaxy as function of redshift. We also suggest mid-IR based selection criteria that can probe different parts of the SFR vs. AGN luminosity diagram.

SFR is usually estimated through a number of indicators such as UV, mid-IR, and total IR luminosities, as well as emission lines (e.g. H$\alpha$) \citep[e.g.][]{Kennicutt1998}. Each of these proxies has its own drawbacks. For example, optical emission lines usually provide reliable SFR estimates (since they probe star-formation on shorter timescales than the integrated UV or IR luminosities) but are much more time-consuming to obtain, especially for high-$z$ sources. On the other hand, UV and mid-IR luminosities can be heavily contaminated by the intrinsic AGN emission and can return biased estimates of the SFR if the AGN emission has not been properly removed.

In our study, we adopt the integrated far-IR luminosity (which mostly traces the cold dust and provides a less biased SFR measure) as our SFR indicator. This task is best accomplished by exploiting the high-sensitivity far-IR observations from the \textit{Herschel} satellite \citep{Pilbratt2010}, which has opened a new window for measuring the rest-frame far-IR properties of QSOs at high-$z$ ($z\gtrsim 1$). Luminous optical QSOs have been extensively studied in the far-IR band and the broadly accepted view is that they tend to live in galaxies with on-going star-formation, at rates similar to those of the star-forming population \citep[e.g.][]{Hatziminaoglou2010,Leipski2013,Rosario2013,Xu2015,Netzer2016,Stanley2017,Hall2018}. Moreover, several studies argue for a positive correlation between the mean SFRs as a function of the bolometric AGN luminosity \citep[e.g.][]{Bonfield2011,Rosario2013,Kalfountzou2014,Gurkan2015,Harris2016,Podigachoski2016,Stanley2017}.
In addition to the far-IR \textit{Herschel} observations, we also include mid-IR data from the WISE satellite. As we will show, combining these multi-wavelength observations can place strong constraints on the SFR of the host galaxy as well as on the AGN power.

The paper is structured as follows. In Sec.~\ref{sec:data} we present the different datasets utilized in the analysis, while the stacking and SED fitting methodology is discussed in Sec.~\ref{sec:methods}. We present our results, compare them with the existing literature, and provide an interpretation in terms of galaxy/SMBH coevolution in Sec.~\ref{sec:results}. Finally, we draw our conclusions in Sec.~\ref{sec:conclusions}.

Throughout this work, we adopt the standard flat cosmology
\citet{Planck2018} with rounded parameter
values: matter density $\Omega_M=0.32$, baryon density $\Omega_b=0.05$,
Hubble constant $H_0=100\, h$ km s$^{−1}$ Mpc$^{−1}$ with $h=0.67$,
and mass variance $\sigma_8=0.81$ on a scale of $8\, h^{−1}$ Mpc.
Reported stellar masses and SFRs (or luminosities) of galaxies
refer to the \citet{Chabrier2003} IMF.

\section{Datasets}
\label{sec:data}
The main focus of this paper is the reconstruction of the mean SED of optical-selected QSOs with the aim of constraining the SFR and the AGN luminosity across cosmic time. To this end, we need access to datasets spanning the electromagnetic spectrum from the optical-UV to the millimeter band. In this section, we first present the public datasets employed in the analysis and then describe the extracted QSO samples.
\subsection{Optical SDSS QSO catalog}
\label{sec:data_sdss}
\begin{figure}[!htbp]
	\includegraphics[width=\columnwidth]{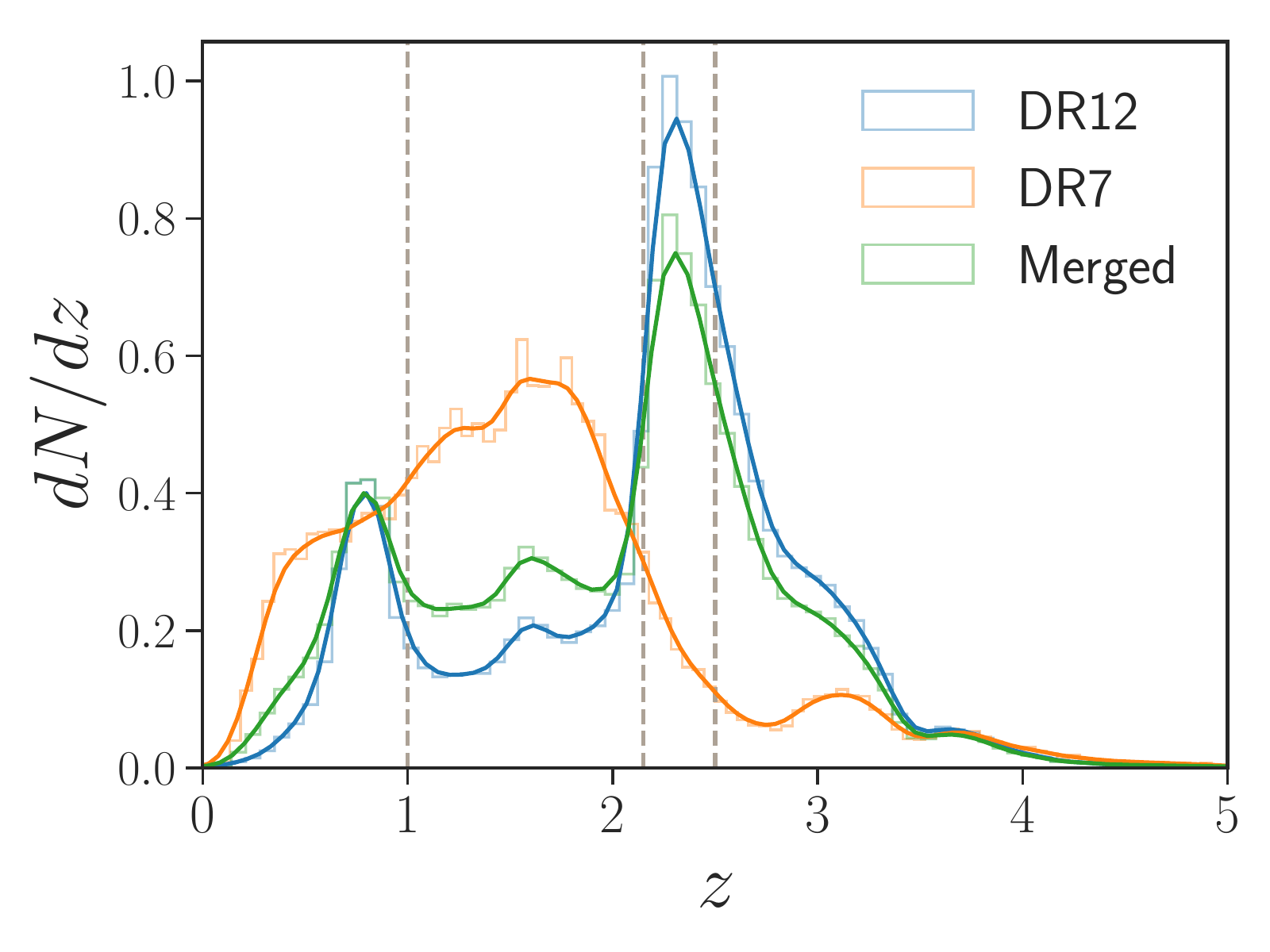}
    \caption{Unit-normalized redshift distribution of the individual SDSS DR7 (orange) and DR12 (blue) QSO catalogs, as well as the merged sample (green) used in this work (after the radio-loud cut has been applied). The vertical dashed lines represent the redshift bins employed in our analysis. The solid smooth lines represent kernel density estimates to the underlying redshift distributions.}
    \label{fig:dNdz}
\end{figure}

We select our QSO sample from the publicly available SDSS-II and SDSS-III Baryon Oscillation Spectroscopic Survey (BOSS) catalogs of spectroscopically confirmed QSOs detected over 9376 $\deg^2$. Specifically, we create a merged sample by combining the QSO catalogs from the seventh \citep[DR7]{schneider:2010}\footnote{Available at \url{http://classic.sdss.org/dr7/products/value_added/qsocat_dr7.html}.} and twelfth SDSS data releases \citep[DR12]{paris:2017}.\footnote{Available at \url{http://www.sdss.org/dr12/algorithms/boss-dr12-quasar-catalog/}.} As can be seen in Fig.~\ref{fig:dNdz}, the DR7 sample mostly contains objects at $z \lesssim 2.5$ while DR12 mainly targets sources at $z \gtrsim 2.15$, although a color degeneracy in the target selection of photometric data causes a fraction of low redshift QSOs to be also observed, resulting in a secondary maximum around $z \sim 0.8$ (see \citet{Ross2012} for a detailed discussion of the QSO target selection process). Approximately 4\% of the DR7 objects at $z>2.15$ have been re-observed for DR12: any QSO present in both DR7 and DR12 is included only once.\footnote{We retain the redundant DR12 information because of the extensive multi-wavelength information available for this catalog.} SDSS QSO catalogs contain several redshift estimates for the sources: in this work we use the SDSS pipeline redshift as our baseline estimate. 
The sources in the merged catalog span a broad range of redshifts, out to $z\sim 7$. We cut the merged catalog to lie in the redshift range $1 \le z \le 5$, discarding QSOs at $z < 1$ and at $z > 5$ because they provide a sparse sampling of the bright QSO population. The resulting catalog contains 313 004 sources (or 386 245 if no redshift cuts are applied). As discussed with greater detail in Sec.~\ref{sec:samples}, the main QSO sample, as well as the other sub-samples considered in this analysis, are extracted from this merged parent QSO catalog. In order to test the evolution of the SED over cosmic time, we further slice the QSO sample into three redshift bins (i) $1 \le z < 2.15$, (ii) $2.15 \le z < 2.5$, and (iii) $2.5 \le  z \le 5$, such that they  approximately contain the same number of sources. 

In this work we make use of the SDSS photometric information, specifically we use the Galactic extinction corrected point spread function (PSF) magnitudes in the $u,g,r,i,$ and $z$ bands which are then converted to fluxes. The DR12 QSO catalog also provides multi-wavelength matching with external surveys. Namely, we use cross-matching to: (i) the latest Faint Images of the Radio Sky at Twenty-Centimeters \citep[FIRST]{Becker1995} radio survey at 1.4 GHz that has almost full overlap with SDSS with a sensitivity of about 1 mJy; (ii) the mid-IR data from the Wide-Field Infrared Survey (WISE), specifically to the newly released AllWISE Source Catalog \citep{2013yCat.2328....0C}; (iii) to the sources detected in the near-IR images from UKIRT Infrared Deep Sky Survey \citep[UKIDSS]{Lawrence2007}. The radio band information from FIRST is exploited to enforce a radio-loud cut in order to minimize the synchrotron emission from the QSOs.

\subsection{Near-IR and mid-IR photometry}
The mid-IR properties of our sample have been defined via cross-matching to the AllWISE sky survey as provided in the SDSS catalogs. WISE has scanned the whole sky in four bands centered at wavelengths of 3.4, 4.6, 12, and 22 $\micron$ ($W1, W2, W3$ and $W4$ channels). WISE All-Sky Source catalog provides PSF-fitting magnitudes and uncertainties in the Vega system for the four mid-IR bands. We convert from Vega magnitudes to flux densities by using zero points of 309.540, 171.787, 31.674 and 8.363 Jy for $W1, W2, W3$ and $W4$ bands respectively. As recommended in the WISE "Cautionary Notes", we only use those matches for which the contamination and confusion flag \texttt{cc\char`_flags} is equal to \texttt{0000}, in order to exclude sources that are flagged as spurious in any band.\footnote{See \url{http://wise2.ipac.caltech.edu/docs/release/allwise/expsup/sec2_2.html\#comp}}. Sources in the catalog with less than a $2\sigma$ significance in a given band are assigned an upper limit given by the integrated flux density measurement plus two times the estimate uncertainty. If the flux density estimate is negative then the upper limit is defined as two times the measurement uncertainty\footnote{See \url{http://wise2.ipac.caltech.edu/docs/release/allwise/expsup/sec2_1a.html}.}. 

Similar to the case of WISE, the near-IR information is included via UKIDSS cross-matching to SDSS QSOs. UKIDSS observes the sky in the four $Y$, $J$, $H$ and $K$ bands (from $\approx 1$ $\mu$m to $\approx 2.2$ $\mu$m) and fluxes are provided in Jy.

\subsection{\textit{Herschel} far-IR photometry and maps}
\label{sec:data_herschel}
\begin{figure}[t]
	\includegraphics[width=\columnwidth]{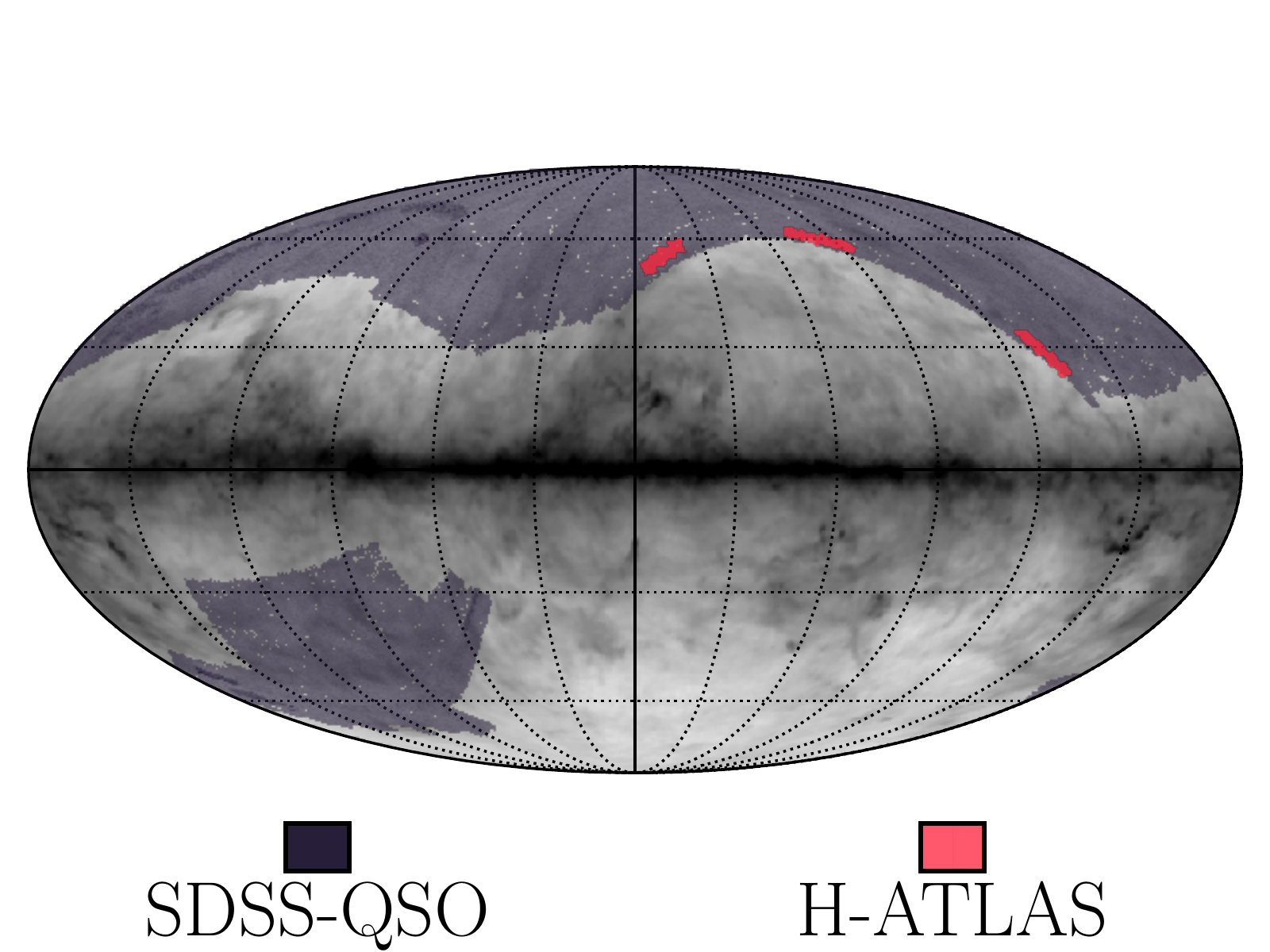}
    \caption{SDSS (purple shaded area) and \textit{Herschel}-ATLAS (pink shaded area) footprints in galactic coordinates. The background image shows the \textit{Planck} thermal dust map from Commander, one of the component separation algorithms adopted by the \textit{Planck} collaboration.}
    \label{fig:maps}
\end{figure}

For our work we make use of data from the first public release of the \textit{Herschel}-Astrophysical Terahertz Large Area Survey \citep[H-ATLAS]{eales:2010}, the largest extragalactic key-project carried out in open time with the \textit{Herschel} Space Observatory. It was allocated 600 hours of observing time and covered about $600\,\hbox{deg}^2$ of sky in five photometric bands with the Photodetector Array Camera and Spectrometer (PACS, 100 and 160 \micron) and the Spectral and Photometric Imaging Receiver (SPIRE, 250, 350 and 500 \micron) instruments. Specifically, we use a subset of the PACS and SPIRE image products provided by the H-ATLAS team\footnote{Available at \url{http://www.h-atlas.org/public_data/}.} \citep[see][ for a detailed discussion of the maps processing and properties]{valiante:2016}, covering the three equatorial fields surveyed by the GAMA spectroscopic survey: GAMA09 (G09), GAMA12 (G12), and GAMA15 (G15), each one of about $\simeq 54\deg^2$. All the maps utilized in our analysis have had the large scale background subtracted through the application of the \textit{Nebuliser} algorithm, so that extended emission from the Galaxy or from any other extra-galactic signal on scales larger than $\approx 3\arcmin$ is removed. In addition, SPIRE maps have also been convolved with a matched-filter derived from the PSF, such that the value in each pixel represents the best estimate of the measured flux that a point source would have in that position. We recall that the SPIRE maps are provided in Jy/beam units while PACS maps are released in Jy/pixel. The pixel sizes are $\approx 3\arcsec, 4\arcsec, 6\arcsec, 8\arcsec, 12 \arcsec$ at 100, 160, 250, 350, $500\,\mu$m respectively, while the angular resolutions at the different frequencies are provided in Tab.~\ref{tab:summary_telescopes}. The three GAMA fields, as well as the SDSS footprint, are shown in Fig.~\ref{fig:maps}. In the case of our baseline QSO sample outlined in Sec.~\ref{sec:data_sdss}, approximately 2300 QSOs fall within the H-ATLAS footprint in each redshift bin.

\subsection{Additional maps: \textit{Planck} and AKARI}
We complement the far-IR information from \textit{Herschel} with \textit{AKARI} data at higher frequencies and with \textit{Planck} observations at similar and lower frequencies. 

The \textit{Planck} satellite has scanned the entire sky in nine frequency bands from 30 to 857 GHz. Its broad spectral coverage allows it to characterize and separate the Galactic and extra-galactic emissions, making it particular suited for astrophysical studies. For a detailed description of the public data release and the scientific results by the \textit{Planck} team, we refer the reader to \citep{planck_overview}. In this work we make use of the publicly available\footnote{\url{http://pla.esac.esa.int/pla/}} 2015  single frequency intensity \textit{Planck} maps from 30 to 857 GHz (see Tab.~\ref{tab:summary_telescopes} for a summary of the maps properties). Maps are provided in the \texttt{HEALPix}\footnote{\url{http://healpix.jpl.nasa.gov}} \citep{healpix} format at a resolution of $N_{\rm side}=2048$, corresponding to $1.7\arcmin$ pixel size.  Maps up to 353 GHz are converted from
units of $K_{\rm CMB}$ to MJy/sr with the unit conversion factors given in the \textit{Planck} 2015 release \citep{planck-compsep2016}. In addition, we utilize \textit{Planck} Cosmic Infrared Background (CIB) maps obtained with the Generalized Needlet Internal Linear Combination method \citep[GNILC]{gnilc}  to estimate the contamination of CIB in our measurements of QSOs fluxes at 353, 545 and 857 GHz. GNILC is a generalization of the internal linear combination (ILC) method capable of extracting multiple correlated sources of emission in a multi-frequency data set taking into account the properties of such signals as function of both direction in the sky and angular scale using needlets. We refer the reader to \cite{remazeilles2011} for further details on the GNILC method.

In order to bridge the observational gap between the mid-IR band covered by WISE and the far-IR portion of the spectrum sampled by \textit{Herschel}, we also make use of the all-sky far-IR data from \textit{AKARI} satellite \citep{doi:2015}. \textit{AKARI} carried out a photometric survey over more than 99\% of the sky in four bands centered at 65, 90, 140, and 160 $\micron$,  with spatial resolutions ranging from 1$\arcmin$ to 1.5$\arcmin$. The absolute flux calibration was performed through comparison of \textit{AKARI} images with COBE/DIRBE data sets. A reliable flux calibration down to low intensities is provided only for the 90 $\micron$ channel, which also has the lowest noise level. For this reason we only consider the 90 $\micron$ \textit{AKARI} map for the present analysis. The map\footnote{Available at \url{http://cade.irap.omp.eu/dokuwiki/doku.php?id=akari}.} is provided in the \texttt{HEALPix} format at a resolution of $N_{\rm side}=4096$, corresponding to $0.86\arcmin$ per pixel, in MJy/sr units.

\begin{center} 
\begin{deluxetable*}{c|ccccccccccc|cccccc|c}
\tabletypesize{\footnotesize}
\tablecaption{Overview of the properties of the maps used in the stacking analysis. The observational wavelength of the \textit{Planck}  satellite have been computed from the effective frequencies of the observational bands \citep{lfi-spectra, hfi-spectra}.\label{tab:summary_telescopes}}
\tablehead{
\colhead{} & \colhead{} & \multicolumn{9}{c}{\textit{Planck}} & \colhead{} & \multicolumn{5}{c}{\textit{Herschel}} & \colhead{} & \colhead{\textit{AKARI}} 
} 
\startdata 
Frequency (GHz)   && 30 & 44 &  70 & 100 & 143 & 217 & 353 & 545 & 857 & & 600 & 857 & 1200 & 1870 & 3000 & & 3300 \\
Wavelength (\micron) && 10600 & 6800 & 4300 & 3000 & 2100 & 1350 & 830 & 540 & 350 & & 500 & 350 & 250 & 160 & 100& & 90 \\
Beam FWHM && 32\arcmin.3  & 27\arcmin.12 & 13\arcmin.31 & 9\arcmin.68   & 7\arcmin.30  & 5\arcmin.02 & 4\arcmin.94 & 4\arcmin.83 & 4\arcmin.64 && 35\arcsec.2 & 24\arcsec.2 & 17\arcsec.8 & 13\arcsec.7 & 11\arcsec.4 && 1\arcmin.30  \\
\enddata
\end{deluxetable*}
\end{center}

\subsection{Sample selections} \label{sec:samples}

In this subsection we describe the construction of our main QSO sample and sub-samples that are used in our analysis.

All of the QSO samples analyzed in this work are extracted from the merged SDSS DR10 and DR12 QSO catalog introduced in Sec.~\ref{sec:data_sdss}. From this parent QSO sample, which contains 386 245 sources, we create three samples by gradually enforcing additional selection criteria:

\begin{itemize}
\item \textit{AllQSO} sample: Our main QSO sample includes 313 004 sources within the redshift range of $1 \le z \le 5$, which we divide in three redshift bins. As mentioned in Sec.~\ref{sec:data_sdss}, we additionally use the radio band information from FIRST to impose a radio-lout cut that minimizes the emission due to synchrotron. Specifically, we reject any object in the parent SDSS QSO catalog that has a counterpart in FIRST, reducing the overall sample size by approximately 4\%. These cuts define our main QSO sample, also referred to as AllQSO in the following figures and tables. The angular distribution of these QSOs is shown as the purple shaded area in Fig.~\ref{fig:maps}. Note that \textit{Herschel} observations only cover a fraction of the whole SDSS footprint. However, unless otherwise stated, when estimating the average QSO SED we combine all the available datasets, i.e. we do not only consider the objects that fall in the H-ATLAS footprint. By doing so, we assume the different patches of the sky to be statistically independent. We test the representativity of the H-ATLAS sub-sample in App.~\ref{sec:HATLAS_subsample_check}.
\item \textit{W4} sample: Most of the optical SDSS QSOs are undetected at an observed wavelength of 22 $\mu$m. In creating this sub-sample, we first start from our main QSO sample (AllQSO) and require a detection in the $W4$ channel of \textit{WISE}, i.e. $S/N \ge 2$. As discussed later, this selection implies a stronger mid-IR emission and, as a consequence, higher emission in the far-IR and UV-optical bands. The number of QSOs selected in this case is approximately 10\% of the main QSO sample and depends on the redshift bin as well as on the specific datasets. For example, in the case of SDSS we find $\sim 13 000$ objects in the first redshift bin and $\sim 8 000$ in the last one, while for \textit{Herschel} the numbers are roughly $\sim 130/200$ for the first/last bin respectively.
\item \textit{W4 + W4/r>100} sample: Since we are also interested in studying the QSOs which live in obscured environments, and considering that the $W4$ selection also increases the mean UV-optical emission (see Fig.~\ref{fig:sed_samples}), we add a further constraint to the W4 sub-sample by requesting individual QSOs with the lowest UV-optical emission. In order to achieve this, we impose a ratio between the mid-IR ($W4$) emission and the optical one ($r$ band) greater than 100 in flux density. The number of SDSS QSOs that fulfills both criteria is around 6100, while the ones falling in the H-ATLAS footprint are approximately 100. If we raise the $S_{W4}/S_r$ threshold to higher values, e.g. $S_{W4}/S_r > 500$, we are left with only a handful of sources, therefore we set the threshold ratio to 100.
\end{itemize}

The three samples outlined above include QSOs which are both undetected (the majority) and detected in the far-IR bands covered by \textit{Herschel}. To assess the differences between the reconstructed SED of sub-mm detected and undetected QSOs, we proceed in the following way. We start from the H-ATLAS catalogs \citep{valiante:2016,Bourne2016} and cross-match them with each one of the three QSO samples described above, retaining only those objects which are detected at \textit{Herschel} frequencies. By doing so, we create three more restrictive QSO samples with the additional property of containing only sub-mm detected QSOs. Note that, differently from the previous case, we limit the analysis the sole QSOs that fall in the H-ATLAS footprint. Moreover, in this case there is no need to perform the stacking analysis on the H-ATLAS maps to infer the sub-mm density flux as the best-fit PACS/SPIRE fluxes are already provided in the H-ATLAS catalogs. Requiring the sub-mm detection and limiting the analysis over the H-ATLAS footprint severely reduce the number of selected QSOs. For this reason, we decided to consider only one redshift bin between $1<z<5$. The number of sources included in these cross-matched samples is 147 for the main QSO sample, 36 for the "\textit{W4}" selection, and 7 in the case of the "\textit{W4 W4/r> 100}" sub-sample. Unfortunately, the number of sources is not sufficient to carry out the stacking analysis with \textit{AKARI} or \textit{Planck}. Tab.~\ref{tab:QSO_numbers} summarizes the number of objects available in the SDSS, WISE, and \textit{Herschel} surveys for all the QSO samples we create.

\begin{table}[]
\caption{The optical-selected quasar samples that we create for this work, along with their redshift coverage. Note that the number of stacked sources depends on the specific survey and wavelength. We report here the number of sources stacked in SDSS, WISE and \textit{Herschel} as representative of the surveys most sensitive to the optical, mid-IR, and starburst components of the SED respectively.}
\label{tab:QSO_numbers}
\resizebox{\columnwidth}{!}{%
\begin{tabular}{ccccc}
\hline
Sample                                                                                   & Redshift bin    & $N_{\rm SDSS}$ & $N_{\rm WISE}$ & $N_{\rm Herschel}$ \\ \hline
\multirow{3}{*}{AllQSO}                                                                  & $1< z < 2.15$   & 62955          & 22868          & 2460               \\
                                                                                         & $2.15< z < 2.5$ & 87644          & 23475          & 2345               \\
                                                                                         & $2.5 < z < 5$   & 92042          & 23391          & 2263               \\ \hline
AllQSO  + H-ATLAS detected                                                               & $1< z < 5$      & 147            & 64             & 147                \\ \hline
\multirow{3}{*}{W4}                                                                      & $1< z < 2.15$   & 13285          & 6322           & 125                \\
                                                                                         & $2.15< z < 2.5$ & 8277           & 3950           & 198                \\
                                                                                         & $2.5 < z < 5$   & 8141           & 3885           & 194                \\ \hline
W4 + H-ATLAS detected                                                                    & $1< z < 5$      & 36             & 34             & 36                 \\ \hline
W4 + W4/r \textgreater 100                                                               & $1< z < 5$      & 6171           & 2974           & 103                \\ \hline
\begin{tabular}[c]{@{}c@{}}W4 + W4/r \textgreater 100 \\ + H-ATLAS detected\end{tabular} & $1< z < 5$      & 7              & 6              & 7                  \\ \hline
\end{tabular}%
}
\end{table}

\section{Methods}
\label{sec:methods}
The main focus of this work is the estimation of the mean SED of optically selected QSOs from the near-ultraviolet to the millimeter band and, through that, constrain the average star-formation and AGN luminosity as function of cosmic time. Since the typical flux density of optically selected QSOs is fainter than \textit{Herschel}, \textit{Planck} and \textit{AKARI} detection limits, we rely on a stacking analysis to recover photometric information for the undetected sources. Stacking is a powerful technique that allows the extraction of signal hidden in noisy datasets, albeit at the price of losing information about the individual sources. In particular, stacking can be shown to be an unbiased maximum likelihood estimator of the average flux density of a catalog, provided that background noise is Gaussian or that the data are confusion-limited \citep{marsden:2009}.
Note that we assume the different patches of the sky to be statistically independent for the purpose of stacking.

\subsection{Herschel stacking}
We perform the stacking analysis in each redshift bin by cutting out $5\arcmin \times 5\arcmin$ stamps at the locations of all the QSOs in our sample, thus creating a pixel cube. The number $N$ of pixels in each stamp then depends on the pixel resolution and as such, it varies between different bands. The mean of the \textit{Herschel} maps is not zero because of the application of the \textit{Nebuliser} algorithm. As recommended by \citet{valiante:2016}, we set to zero the mean of each map before performing the stacking analysis. After extracting the cutouts at the different redshift bins and wavelengths, we combine them into a single image, a process known as data cube reduction. In literature this is usually achieved by computing the mean or the median flux of all the stamps in a given pixel. In this work we use the mean, in order to represent the typical flux density of the sample.
To this end, we use the noise maps extracted at the same QSO locations to perform a weighted average of each pixels in the stacked image:
\begin{equation}
\bar{S}_{\nu} = \frac{\sum_i^{N_{\rm QSO}}w^{i}_{\nu} S^{i}_{\nu}}{\sum_i^{N_{\rm QSO}}w^{i}_{\nu}}.
\end{equation}
Here, $S^i_{\nu}$ is the pixel flux density in the stamp of the QSO $i$ at a band $\nu$, $w^i$ is the inverted noise flux density in the same pixel, and the sum runs over all the $N_{\rm QSO}$ sources in a given redshift bin. 
We adopt two different approaches to measure the stacked flux from the PACS and SPIRE cutouts. For the PACS stacks, which are in units of Jy/pixel, we perform aperture photometry and integrate the flux density out to 5.6$\arcsec$ and 10.2$\arcsec$ for the 100 and 160 $\micron$ channels respectively. The sky background is calculated in a ring of inner and outer radii of $35\arcsec$ and $45\arcsec$ and subtracted from the previous estimate. In the case of the SPIRE observations, which are in the units of Jy/beam, we fit a 2D Gaussian (plus a constant background) to the image stack and take its peak value as the flux density estimate.

\begin{figure*}[!htbp]
\includegraphics[width=\textwidth]{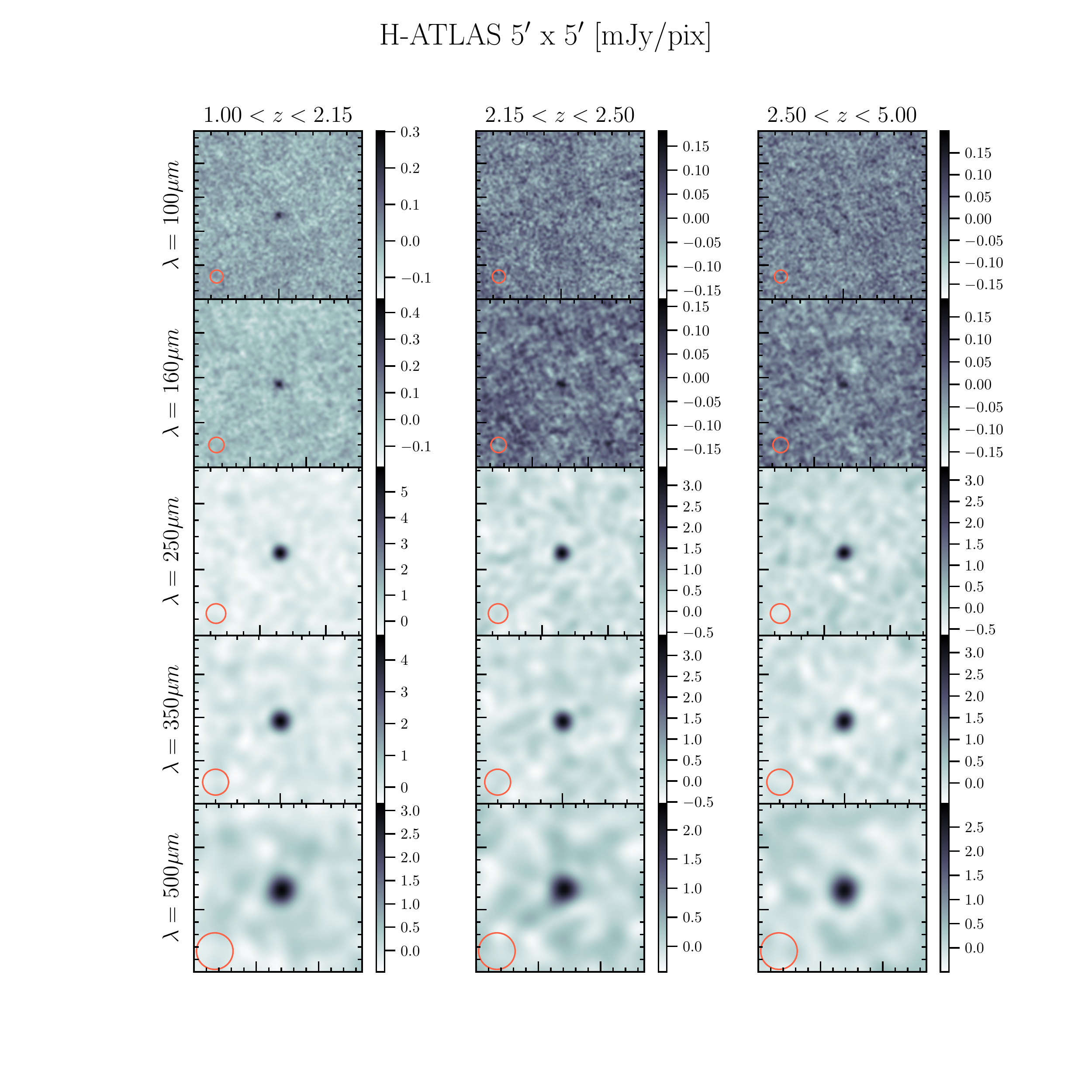}
    \caption{Stacked thumbnail images from H-ATLAS PACS and SPIRE data at the locations of sources included the main QSO sample (AllQSO). The number of objects stacked in each redshift bin is approximately 2300. Each box is $5 \arcmin \times 5 \arcmin$ and the units are mJy/pixel. The orange circle in each panel represents the instrument PSF at the various bands. Its diameter corresponds to the FWHM size.}
    \label{fig:herschel_stamps}
\end{figure*}

We employ a bootstrap approach to estimate the $1\sigma$ errors on the stacked flux in both the PACS and SPIRE bands for each redshift bin. To this end, we randomly re-sample (with replacement) the sources and perform the stacking procedure 100 times for each redshift bin and frequency band. The uncertainty in the flux is then computed as the standard deviation of the measured stacked mean flux in these 100 trials. An example of stacked cutouts in different redshift bins and PACS/SPIRE frequencies at the positions of our main QSO sample is shown in Fig.~\ref{fig:herschel_stamps}.

\subsection{Planck and AKARI stacking}
\textit{Planck} and \textit{AKARI} data sets differ from the \textit{Herschel} maps in that they are full-sky and have a lower angular resolution. For these reasons, as well as to avoid projection effects associated to the conversion from the \texttt{HEALPix} full-sky to the flat-sky geometry and pixelization, we follow \citet{soergel:2017} and rely on a different approach based on matched filtering to extract the QSOs emission in the \textit{Planck} and \textit{AKARI} maps. Matched filtering, first proposed by \citet{matched_filter} in the context of kinetic Sunyaev-Zel'dovich (kSZ) studies, is a powerful technique that allows the construction of an optimal spatial filter to recover the signal of interest (the QSOs flux density in our case) in the presence of a noisy background (the primary CMB, foregrounds, and instrumental noise). One of the assumptions at the base of the matched filter approach is that the spatial signature of the sought-after signal is known. Given that the QSO host halos are unresolved at all \textit{Planck} and \textit{AKARI} frequencies (100 Kpc at $z\sim 2$ subtend an angle of about $0.2\arcmin$, much smaller than their respective angular resolutions, see Tab.~\ref{tab:summary_telescopes}), we can safely assume the QSOs spatial profile to be approximated by the instrumental beam profile. Note that in this way the extracted fluxes are model-independent.

The filtered map harmonic coefficients $a_{\ell m}^{\rm filt}$ are related to the unfiltered ones through $a_{\ell m}^{\rm filt}= F_{\ell}a_{\ell m}$ \citep{schaefer:2006}, where the filter function reads as
\begin{equation}
F_{\ell} = \left [\sum_{\ell} \frac{(2\ell+1)\tilde{B}_{\ell}^2}{4\pi C_{\ell}^{\rm tot}} \right ]^{-1} \frac{\tilde{B}_{\ell}}{C_{\ell}^{\rm tot}}.
\end{equation}
\newline
In the above equation, $\tilde{B}_{\ell}=2\pi\sigma B_{\ell}p_{\ell}$, $B_{\ell}$ is the instrumental beam window function, $p_{\ell}$ is the pixel window function, and $\sigma=\theta_{\rm FWHM}/\sqrt{8\ln 2}$ is a factor that accounts for the beam normalization to unity in real space. Instrumental beams are assumed to be Gaussian with FWHM shown in Tab.~\ref{tab:summary_telescopes}. We checked that this approximation was sufficiently accurate by repeating the analysis for the  \textit{Planck}  frequencies while adopting the $B_{\ell}$ of the \textit{Planck}  Reduced Instrument Model\footnote{\url{https://wiki.cosmos.esa.int/planckpla2015/index.php/The_RIMO}} and finding no significant difference with our baseline results. For the GNILC maps we assumed a Gaussian beam with 5$\arcmin$ FWHM following \cite{gnilc}.
The power spectrum describing all the noise sources is taken to be the total power spectrum of the unfiltered maps, $C_{\ell}^{\rm tot}$, since the QSOs emission signal is negligible with respect to primary CMB, foreground, and instrumental noise. The total power spectrum is extracted from the full-sky maps using the \textsc{Xpol} code \footnote{\url{https://gitlab.in2p3.fr/tristram/Xpol}} \citep{tristram2005} after applying an apodized ($2^{\circ}$) 40\% Galactic mask to mitigate the effect of Galactic foregrounds and a point source mask obtained by the combination of the \textit{Planck} 2015 Catalog of Compact Sources and the \textit{AKARI} Far-Infrared Surveyor Bright Source Catalog. As a conservative choice to mitigate any residual large scale foreground and following \cite{soergel:2017}, we further apply an additional smooth high-pass filter to remove large angular scales at $\ell\lesssim 300$.
After filtering the full-sky maps, we can estimate the flux density for the QSO at position $i$ by multiplying the pixel value $I_{\nu}^{\rm filt}(\nver_i)$ of the filtered map by the beam area,
\begin{equation}
S_{\nu}^{i} = I_{\nu}^{\rm filt}(\nver_i) \int \textrm{d}\Omega\, B_{\nu}(\theta).
\end{equation}
Then, the mean flux density is computed as 
\begin{equation}
\bar{S}_{\nu} = \frac{1}{N_{\rm QSO}}\sum_i S_{\nu}^{i},
\end{equation}
while the frequency-frequency covariance reads as 
\begin{equation}
C_{\nu\nu^{\prime}} = \frac{1}{N_{\rm QSO}(N_{\rm QSO}-1)}\sum_i (S_{\nu}^{i}-\bar{S}_{\nu})(S_{\nu^{\prime}}^{i}-\bar{S}_{\nu^{\prime}}),
\end{equation}
from which we extract the $1\sigma$ error bars as the square root of the 
diagonal.

\subsection{ Mean flux densities from catalogued sources}
We also estimate the QSOs fluxes at the catalog level using the available SDSS, UKIDSS, and WISE photometry. In order to be consistent with the stacking analysis discussed above, we report the mean flux density as the representative quantity for the QSO sample at the different bands. For the SDSS and UKIDSS fluxes we estimate the $1\sigma$ uncertainties as the standard error deviation of the flux density distribution of the individual sources divided by the square root of the number of objects. In the case of WISE photometry, the simple mean of the QSOs flux distribution would yield a biased value since the fraction of sources with upper limits in the $W3$ and $W4$ channels range from 32\% to 44\% and from 70\% to 81\% in the three redshift bins respectively (for our baseline QSO sample). Given that the upper limits show a random enough distribution, we exploit the non-parametric Kaplan-Meier estimator \citep{Feigelson1985}
to take into account the upper limits and calculate the WISE mean flux density. To assess the WISE uncertainties we conservatively consider two extreme cases, one where we calculate the mean flux assuming all the sources to be detected (upper error bar), and one where we compute the mean flux assigning a null flux to the sources for which only upper limits are provided (lower error bar). 

\subsection{Modeling the total SED}
\label{sec:sed_model}
In order to study the average physical properties of our selected QSOs we need to perform a SED decomposition. We are mainly interested in deriving the bolometric luminosity for each component and comparing their variation with redshift and for the different sub-samples described in the next sub-section. Therefore, a detailed analysis of the $0.5-500 \mu m$ SEDs is beyond the purpose of this paper. We simplified the SED emission in three components: AGN emission from the central engine of the QSOs, a starburst emission from the star formation activity in the galaxy host, and a mid-IR emission from the torus surrounding the central engine. 
\begin{enumerate}
\item \textit{UV-optical component}: To represent the UV-optical emission (residual after obscuration) from normal quasars we selected the AGN intrinsic template based on \citet{Elvis1994} developed by \citet[see their Appendix C]{Xu2015}. We assume that this SED remains unchanged for all the considered sub-samples and redshifts.
\item \textit{Starburst component}: We used the \citet{Pearson2013} two gray-body template that was found to best describe the intermediate-high redshift starburst galaxies observed by \textit{Herschel}. Although simplistic, it provides enough information to perform a rough estimation of a photometric redshift and to fit the stacked far-IR emission (see Sec.~\ref{sec:samples}).
\item \textit{Torus component}: Classical SED decomposition approaches to mid-IR observations, i.e. torus emission, use rigid templates that do not provide a good fit to our stacked QSO SEDs \citep[e.g.][]{Richards2006,Elvis1994,Lani2017}. Following \citet{Lyu2017}, we decided to fit the mid-IR emission by taking into account three gray-body components in the following temperature ranges: 1000-2000 K, 700-1000 K and 200-700 K. For all of them we fixed the dust emissivity to $\beta=1.5$ as found in studies of high-$z$ AGN \citep[e.g.][]{Beelen2006,Lyu2017}. 
\end{enumerate}

Taking into account these three components we perform a two-steps SED decomposition. i) We exclude the mid-IR data ($2 \mu m < \lambda < 90 \mu m$) focusing only on the optical ($\lambda < 2 \mu m$) and far-IR data ($\lambda > 90 \mu m$). We fit at the same time the UV-optical and starburst components, i.e. we determine the normalization (vertical shift) for each component and the common redshift (horizontal shift or a kind of mean photometric redshift). The \texttt{curve\_fit}, a non-linear least squares algorithm from \texttt{scipy} package, allows us to derive the best-fit parameters for the sum of both components to the measured data. The estimated photometric redshifts are very close to the spectroscopic mean redshift in each bin as expected. ii) We add the mid-IR measurements and we attempt to fit them with the addition of a Torus component described as three gray-body functions. We start with the UV-optical and starburst component considering the best-fit parameters determined in the previous step. The best-fit parameters from the previous step remains fixed at this step. Then we sum the three grey-body functions at the same best-fit redshift from the previous step. Finally, using again \texttt{curve\_fit}, we determine the best-fit parameters for the three grey-body functions.

\begin{figure*}[!htbp]
\includegraphics[width=\textwidth]{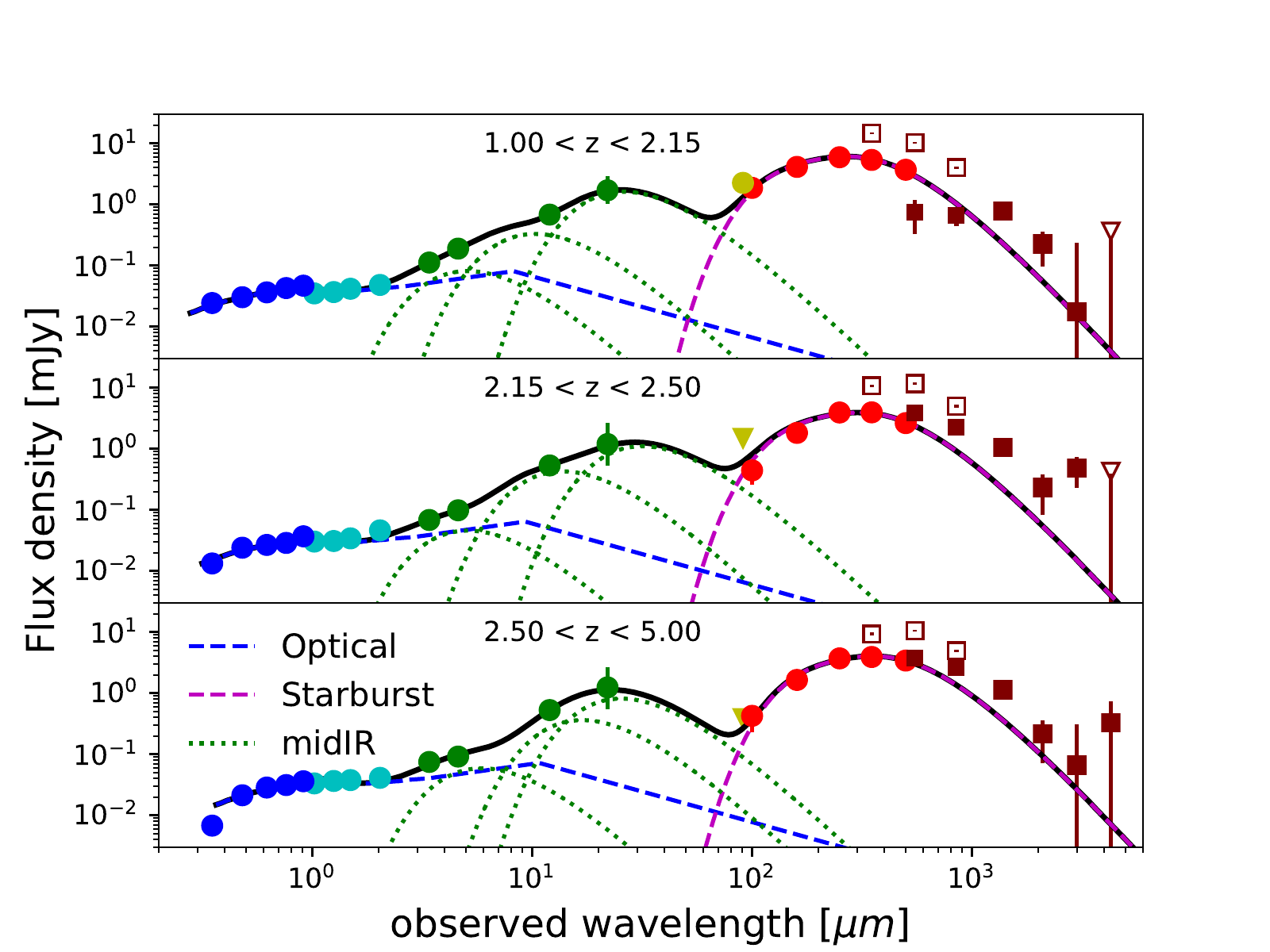}
    \caption{Averaged SEDs obtained for the main sample of QSOs in the three redshift bins with the following photometry: SDSS (dark blue), UKIDSS (light blue), \textit{WISE} (green), \textit{AKARI} (yellow), \textit{Herschel} (SPIRE + PACS, red) and \textit{Planck} (dark red). The CIB contamination estimated with the GNILC map has been subtracted from the average flux density at 540 and 830 $\mu m$. Empty markers at those frequencies represent the values of the measurements prior to this correction. CIB-subtracted measurements at 350 $\mu m$  are not shown as they are affected by systematics in the GNILC maps.  In the case of \textit{AKARI} and \textit{Planck}, lower triangles indicate upper limits. As described in detail in the text, the SEDs are modelled using three components: UV-optical emission \citep[dashed blue]{Xu2015}, torus emission (dotted green, three gray-body components) and a starburst emission \citep[dashed magenta]{Pearson2013}. Note that the \textit{Planck} data (empty/filled dark red squares) were not used in the fitting process.}
    \label{fig:sed_fit}
\end{figure*}

Obviously, the mid-IR data fit is degenerate: six free parameters  (three temperatures plus three normalization values) versus 4 photometric points. However, as the UV-optical and starburst components are already fixed and the temperatures are restricted to a narrow range, the algorithm is able to converge to a reasonable fit, at least, good enough to estimate the mid-IR bolometric luminosity. Consequently, the estimated temperatures are reasonable but very uncertain and, for this reason, they are not presented nor discussed here. We plan to study in detail the mid-IR range in a future work.  See Fig.~\ref{fig:sed_fit} for an example of the SED fitting for the main QSO sample in the three redshift bins. Note that the \textit{Planck} data (empty/filled dark red squares) are just plotted for a visual comparison but not used in the fitting process. As commented later on, we defer the analysis of CMB and radio frequencies to future work.

Once the fit is obtained, we estimate three bolometric luminosities, one for each of the components: for the AGN component we consider only the UV-optical SED, for the mid-IR the sum of the three gray-body fitted functions and for the far-IR the starburst SED. This process is repeated for each sample and redshift bin, as discussed in Sec.~\ref{sec:reconstructed_sed}.

As a robustness test, we check that our derived averaged SEDs are independent of the exact applied methodology (see Appendix \ref{sec:sed_test} for more details).

\section{ Results and Discussion}
\label{sec:results}
In this section we discuss the main findings of our analysis. We start by showing the reconstructed SED for the different QSO samples and reporting the inferred bolometric luminosities of the various QSO/host-galaxy components. We then compare our mean SED with those present in literature and conclude by discussing the results in terms of the galaxy-SMBH coevolutionary framework.

\subsection{Reconstructed SEDs}
\label{sec:reconstructed_sed}
In Fig.~\ref{fig:sed_fit} we present the measured average SEDs for our main QSO sample in the three redshift bins (see Table \ref{tab:QSO_luminosities} for their estimated photometric redshifts), along with the best-fit SEDs and their decomposition in terms of the single components described in Sec.~\ref{sec:sed_model}. In all three cases we can easily identify the UV-optical emission associated to the AGN, the starburst component at sub-mm wavelengths as well as the mid-IR bump sourced by the heated torus emission. 

Note also that although there is a qualitative good agreement between the extrapolated SEDs and \textit{Planck}'s measured fluxes at mm and radio wavelengths, we do not make use \textit{Planck} data in the fitting process as the flux derived from the stacking appears to generally overestimate the flux density. The overestimation of the flux density becomes more important at the shorter wavelengths of \textit{Planck}, in particular at those overlapping with the \textit{Herschel} bands, i.e. $540$ and $830\, \mu$m. Residual miscalibration between the \textit{Herschel} and \textit{Planck}  cannot account for the difference in the measured flux density. Thus, we suspect that the coarser resolution and higher noise levels might lead to a measured flux density that systematically picks up background gray-body-like signal not associated with the QSOs, e.g. CIB or residual Galactic dust. If we subtract the estimated CIB contamination obtained from the stacking of the CIB GNILC maps at  540, 830 $\mu$m we recover values of the flux densities compatible with the expected values of our best-fit SED. However, the GNILC maps at $340\, \mu m$ are highly dominated by spurious systematic effects. The estimated flux density of this map, in fact, exceeds the value extracted from the \textit{Planck} single wavelength map, which is an unphysical situation. Hints that the GNILC CIB flux density might be overestimated can also be observed in the $ 1< z < 2.25$ redshift bin at $540 \, \mu$m. Despite templates of the Galactic dust emission have been published by the \textit{Planck}  collaboration using several component separation methods, we did not perform stacking using these maps as on small angular scales they are heavily contaminated by CIB emission \citep{gnilc, planck-compsep2016}. 
For all the reason just discussed, we decided to postpone an accurate inclusion of \textit{Planck} data to a future work. We stress that our findings also suggest that any use of the \textit{Planck} data for analysis similar to ours should be treated with caution, in particularly in the highest frequency channels.

\begin{figure*}[!htbp]
\includegraphics[width=\textwidth]{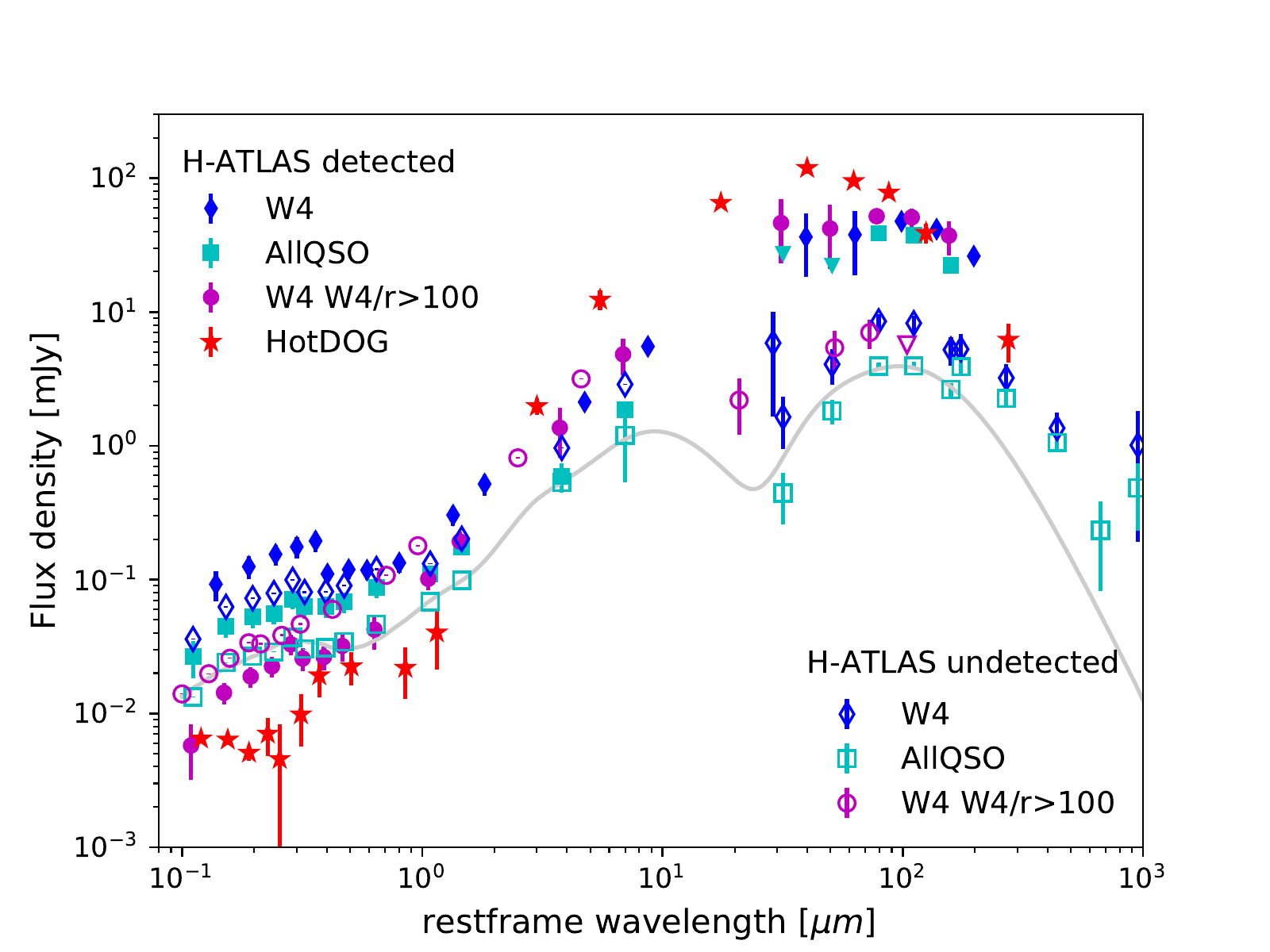}
    \caption{Averaged SEDs obtained for the different samples of QSOs in the redshift range $2.15 <z < 2.5$ (or $1<z<5$ for the H-ATLAS detected cases, see text for more details). Filled symbols correspond to SEDs where only sources detected in H-ATLAS catalogs are considered (no map-based stacking was needed in this case). On the contrary, the SEDs estimated considering also undetected H-ATLAS sources (stacking applied) are represented with empty symbols. Lower triangles indicate upper limits.}
    \label{fig:sed_samples}
\end{figure*}

The estimated mean SEDs for all the samples described in Sec.~\ref{sec:samples} are shown and compared in Fig.~\ref{fig:sed_samples}. We remind that the SEDs for the $W4 + W4/r > 100$ sample and all of the H-ATLAS detected samples are calculated over the redshift range $1 < z < 5$, while the remaining SEDs are obtained for the intermediate redshift bin $2.15 < z < 2.5$.
When looking at the samples comprising both undetected and detected \textit{Herschel} (empty symbols) we notice that requiring a $S/N \ge 2$ in the $W4$ band results in an overall shift to higher flux densities with respect to the values found for the main QSO sample. Instead, imposing a further constraint on the mid-IR to optical flux ratio $S_{W4}/S_r > 100$ translates to similar or slightly smaller UV-optical flux densities and higher flux values at  mid- and far-IR wavelengths when compared to the baseline QSO sample, while pushing down the measured fluxes at $\lambda \lesssim$ few $\mu$m with respect to the "$W4$" case. As expected, requiring a detection at far-IR frequencies affects the reconstructed SED by boosting the far-IR emission from few mJy to few tens of mJy, almost independently on the specific selection criteria (filled symbols). A similar trend is seen at wavelengths $\lambda \lesssim 100\, \mu$m, where the average fluxes for the sub-mm detected sources are enhanced, although in a negligible way for $W4 + S_{W4}/S_r > 100$ selected sources in the UV-optical band range. In Fig.~\ref{fig:sed_samples} we also plot the SED of the only hot dust-obscured galaxy (HotDOG) from \citet{Fan2016} which is included in our SDSS-based QSO catalog (SDSS 022052.11+013711.1). This object, situated at a redshift of $z=3.138$, is a violently star-forming galaxy featuring a high-temperature dust component with a strong sub-mm emission and is characterized by a $W4$ to $r$ ratio of about 1000. 

In Table~\ref{tab:QSO_luminosities} we report the estimated photometric redshift and the inferred bolometric luminosities obtained by applying the fitting method outlined in Sec.~\ref{sec:sed_model} to each SEDs. As can be seen, the luminosities associated to the three QSO components tend to increase over cosmic time, with similar UV-optical and mid-IR luminosities (of course when $S_{W4}/S_r$ criterion is not enforced). Once again, as expected, the sub-mm detection criterion results in starburst luminosities above $\log_{10}{(L_{\rm SB}/L_{\odot})} \gtrsim 12.60$. 

\begin{table*}[t]
\centering
\caption{Inferred bolometric luminosities of the optical, mid-IR, and starburst components and estimated photometric redshift for the different samples and redshift bins.}
\label{tab:QSO_luminosities}
\begin{tabular}{ccccccccc}
\hline
Sample & Redshift bin & Photometric $z$  & $\log_{10}{L_{\rm opt}} \, [L_{\odot}]$ & $\log_{10}{L_{\rm mIR}} \, [L_{\odot}]$ & $\log_{10}{L_{\rm SB}} \, [L_{\odot}]$ \\ \hline
\multirow{3}{*}{AllQSO}                                                                  & $1< z < 2.15$ & $1.79^{+0.01}_{-0.01}$  & $12.32^{+0.01}_{-0.01}$                 & $12.30^{+0.17}_{-0.14}$                 & $11.84^{+0.02}_{-0.02}$                \\
                                                                                         & $2.15< z < 2.5$& $2.15^{+0.01}_{-0.01}$ & $12.36^{+0.01}_{-0.01}$                 & $12.34^{+0.22}_{-0.15}$                 & $11.79^{+0.02}_{-0.02}$                \\
                                                                                         & $2.5 < z < 5$ & $2.65^{+0.01}_{-0.01}$  & $12.57^{+0.01}_{-0.01}$                 & $12.52^{+0.24}_{-0.20}$                 & $11.96^{+0.02}_{-0.02}$                \\ \hline
AllQSO  + H-ATLAS detected                                                               & $1< z < 5$   & $1.59^{+0.07}_{-0.07}$   & $12.52^{+0.05}_{-0.06}$                 & $12.40^{+0.17}_{-0.07}$                 & $12.60^{+0.05}_{-0.05}$                \\ \hline
\multirow{3}{*}{W4}                                                                      & $1< z < 2.15$ & $1.68^{+0.02}_{-0.02}$  & $12.61^{+0.01}_{-0.01}$                 & $12.57^{+0.01}_{-0.02}$                 & $12.02^{+0.05}_{-0.06}$                \\
                                                                                         & $2.15< z < 2.5$& $2.15^{+0.03}_{-0.03}$ & $12.79^{+0.01}_{-0.01}$                 & $12.79^{+0.01}_{-0.01}$                 & $12.11^{+0.05}_{-0.05}$                \\
                                                                                         & $2.5 < z < 5$ & $2.65^{+0.03}_{-0.03}$  & $12.99^{+0.01}_{-0.01}$                 & $13.03^{+0.01}_{-0.01}$                 & $12.32^{+0.04}_{-0.05}$                \\ \hline
W4 + H-ATLAS detected                                                                    & $1< z < 5$   & $1.53^{+0.14}_{-0.14}$   & $12.72^{+0.10}_{-0.11}$                 & $12.64^{+0.11}_{-0.12}$                 & $12.62^{+0.09}_{-0.10}$                \\ \hline
W4 + W4/r \textgreater 100                                                               & $1< z < 5$   & $2.21^{+0.27}_{-0.27}$   & $12.83^{+0.01}_{-0.01}$                 & $13.29^{+0.01}_{-0.01}$                 & $12.51^{+0.08}_{-0.10}$                \\ \hline
\begin{tabular}[c]{@{}c@{}}W4 + W4/r \textgreater 100 \\ + H-ATLAS detected\end{tabular} & $1< z < 5$    & $3.80^{+0.04}_{-0.04}$  & $12.27^{+0.12}_{-0.14}$                 & $13.49^{+0.19}_{-0.23}$                 & $12.95^{+0.12}_{-0.13}$                \\ \hline
\begin{tabular}[c]{@{}c@{}}HotDOG \\ SDSS 022052.11+013711.1\end{tabular}                & z = 3.138 & ---   & $11.02^{+0.08}_{-0.09}$                 & $12.42^{+0.16}_{-0.19}$                 & $12.63^{+0.07}_{-0.08}$                \\ \hline
\end{tabular}
\end{table*}

\subsection{Comparison with SEDs from literature}
\label{sec:comparison_sed}
\begin{figure*}[!htbp]
\includegraphics[width=\textwidth]{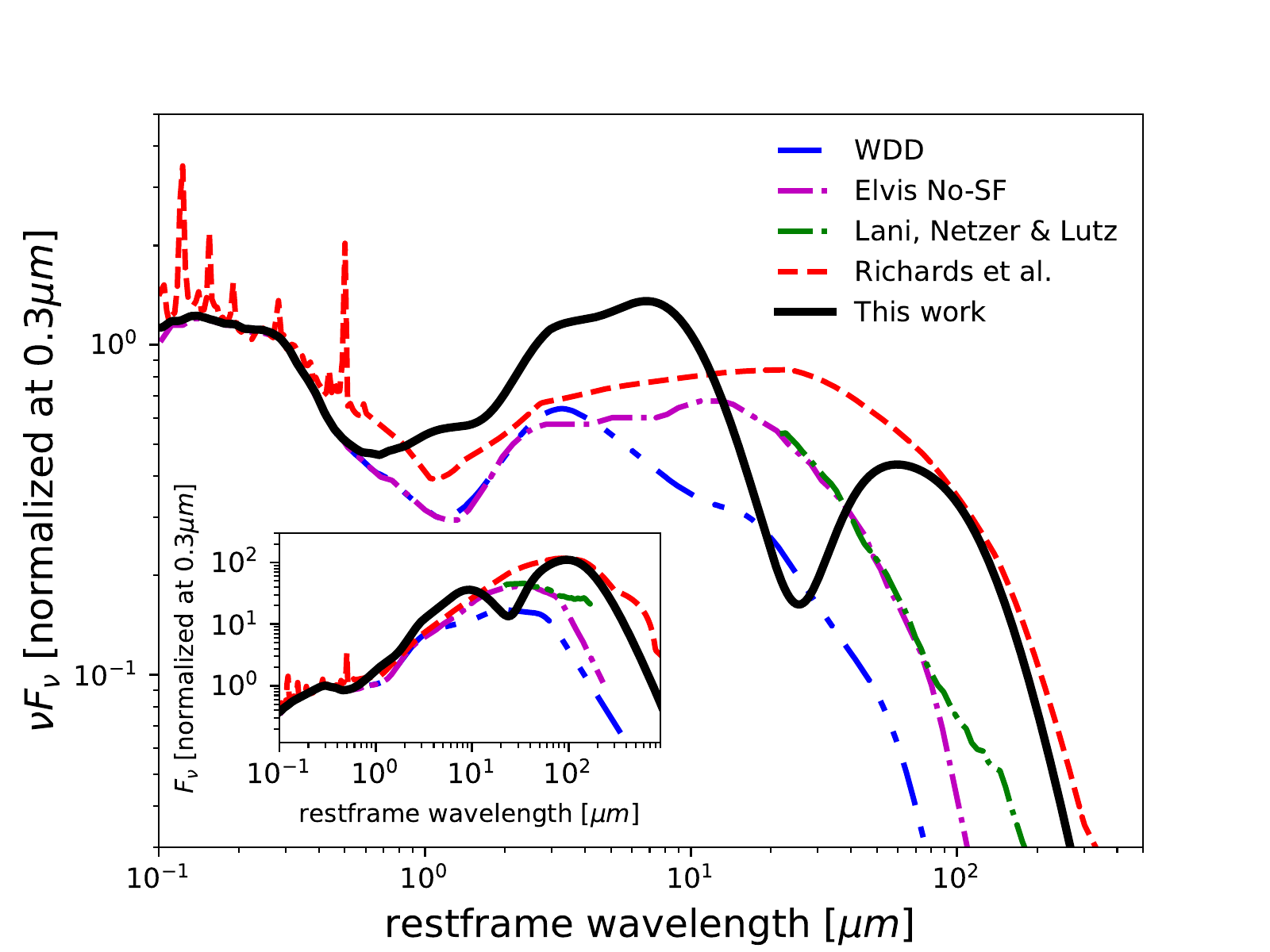}
    \caption{Comparison of the mean SED reconstructed in this work in the redshift range $2.15 < z < 2.5$ (solid black line) with classic templates from literature. We show the SED from \citet{Richards2006} obtained using SDSS-selected QSOs in combination with \textit{Spitzer} data (dashed red line), the template by \citet{Elvis1994} as revisited by \citet{Xu2015} from star formation in the host galaxy (dot-dashed magenta), and the WDD template by \citet{Lyu2017}. Moreover, we have also plotted the most recent intrinsic AGN SED derived by \citet[green line]{Lani2017}.}
    \label{fig:sed_samples2}
\end{figure*}

In Fig.~\ref{fig:sed_samples2} we illustrate the stacked rest-frame SED for the full
QSO sample from our analysis (in the redshift range
$2.15<z<2.5$), and compare it to classic AGN templates from the
literature. Specifically, we consider: the template by \citet[red dashed line]{Richards2006}  obtained from a sample of 259 SDSS-selected quasar with \textsl{Spitzer} data; the template by \citet[magenta dot-dashed line]{Elvis1994} based on a Palomar-Green (PG) sample of radio-quiet QSOs, as revisited by \citet{Xu2015}
from star formation in the host.  
Moreover, we have also plotted the most recent warm-dust deficient template
by \citet[cyan triple-dot-dashed line]{Lyu2017} applying to
high-luminosity PG QSOs and the intrinsic AGN SED derived by \citet[green line]{Lani2017}. This latest SED corresponds to the median SED after subtracting the star formation contribution using a PAH-based method to the PG QSO sample.
All the templates have been normalized
to unity at a rest-frame wavelength of $0.3\,\micron$.

As expected, our SED is very similar in shape to the other templates in the optical/UV part of the spectrum. At around $1-2\, \mu$m it shows a little excess with respect to the other templates, likely due to the combination of the emissions from old stars (e.g., RGB/AGB) in
the host galaxy and from dust near the sublimation temperature ($T\sim 2000$ K). Around $4-6\, \mu$m our SED features a prominent
bump, followed by an abrupt decrease out to $10\, \mu$m. This articulated behavior is not present in the Richards template, while it is just hinted in the Elvis No-SF a WDD templates. In fact,
the stacked data from WISE and \textit{AKARI} are fundamental in revealing
the bump and the decrease in the mid-IR; similar evidences have
also been found in individual hyperluminous quasars and HotDOGs
observed with WISE by \citet[cf. their Fig.~3]{Tsai2015}.
Finally, above $20\,\mu$m our stacked SED increases again and
behaves similarly to the Richards template in the far-IR.

We stress that our SED has been obtained by stacking of an
unprecedented number of quasars with multi-band data from the UV to
the far-IR band (including SDSS, UKIDSS, WISE and \textit{Herschel}). This
has enabled us to reveal detailed features of the average AGN SED
that were missed by previous studies based on more limited
statistics and spectral coverage. In particular, the presence of
the bump around $4-6\, \mu$m and of the steep decrease in the
mid-IR constitute the most relevant finding from our stacked
analysis.

A qualitative physical interpretation of these features may be related to 
the impact of the QSO emission on the surrounding galactic environment. 
Recall that the stacked signal in our
analysis is dominated by individual quasar hosts with far-IR
emission undetected in \textit{Herschel}; the stacked sample 
includes both galaxies with intrinsic SFR below the \textit{Herschel} detection limit, and galaxies with large-scale star formation already substantially reduced 
by the energy/momentum feedback from the QSO. Furthermore, 
a determination of the stellar mass for these sources is fundamental in interpreting the data.

On smaller scales, the dusty torus around the central SMBH, in general
responsible for most of the mid-IR emission, may be
affected by the QSO feedback. 
Specifically, the torus may be partly peeled off, the associated warm-dust covering factor may be
reduced substantially, and the residual dust component may be
instead excited to high temperatures. 
A partially eroded warm-dust torus with a residual hot dust component may indeed be at the
origin of the mid-IR behavior characterizing our measured SED (see the WDD template in Fig. 6). 
However, a detailed radiative transfer model would be required to 
test such an hypothesis; this is beyond the scope of the present work and
will be performed in a forthcoming paper.

\subsection{Interpretation in terms of galaxy-BH coevolution scenario}
\label{sec:interpretation}

\begin{figure*}[!ht]
\includegraphics[width=\textwidth]{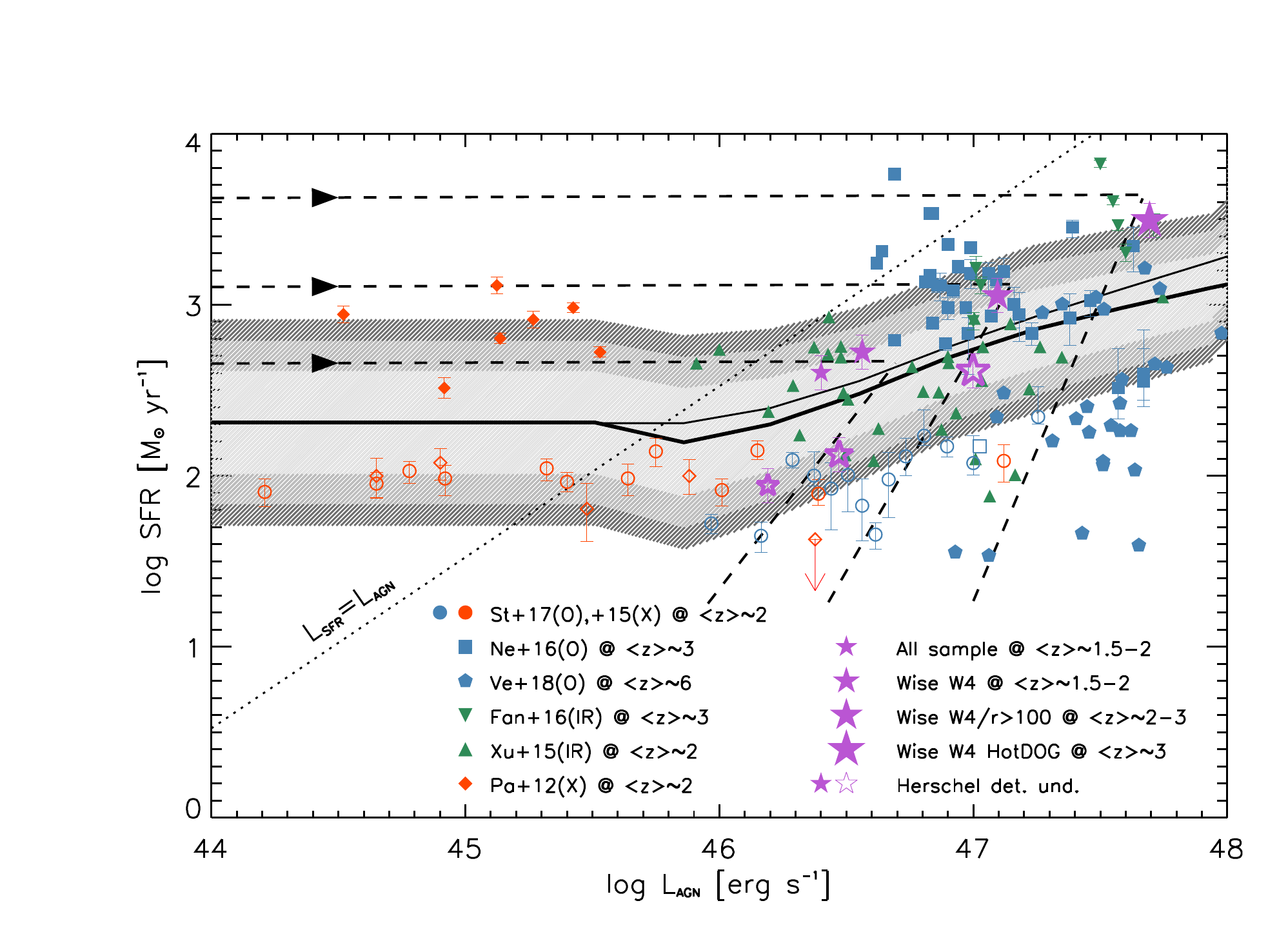}
\caption{SFR vs. $L_{AGN}$ plane between bolometric AGN luminosity and star formation rate in the host galaxy at $z\ga 2$. Dashed lines illustrate three typical evolutionary tracks (forward time direction indicated by arrows) from the in-situ coevolution scenario by \citet{Lapi2014,Lapi2017} (see also \citet{Mancuso2016,Mancuso2017}) corresponding to SFR $\sim 400$, $1200$, and $4000\, M_\odot$ yr$^{-1}$. Dotted line represents the locus where the bolometric luminosity from AGN emission and from star formation are equal; galaxies to the right of such a line are prone to be affected by substantial AGN activity (see text). The thick solid line shows the average relationship and the shaded areas render its $1\sigma$ (light grey), $2\sigma$ (grey), and $3\sigma$ (dark grey) variance, computed as in \citet{Mancuso2016} taking into account number density of galaxies and AGNs, the relative time spent by individual objects in different portions of the tracks, and a far-IR detection threshold around SFR $\sim 100\, M_\odot$ yr$^{-1}$. Literature data are from \citet[circles]{Stanley2015,Stanley2017}, \citet[squares]{Netzer2016}, \citet[pentagons]{Venemans2018}, \citet[inverse triangles]{Fan2016}, \citet[triangles]{Xu2015} and \citet[diamonds]{Page2012}. Red symbols refer to X-ray selected AGNs, blue to optically-selected AGNs, and green to mid-IR selected AGNs; the SFR measurements are obtained from far-IR observations with \textsl{Herschel}, both via detections (filled symbols) and via stacking (empty symbols); the sample by \citet{Venemans2018} is based on detections from ALMA millimetric data. Results from this work are illustrated as magenta stars for the overall sample at $\langle z\rangle\sim 1.5-2$ (tiny stars), for \textsl{WISE} $W4$-detected galaxies at $\langle z\rangle\sim 1.5-2$ (small stars), for \textsl{WISE} $W4$-detected galaxies with $W4/r$-band luminosity ratio exceeding $100$ at $\langle z\rangle\sim 2-3$ (medium-size stars), and for the unique HotDOG falling in the \textsl{Herschel}-ATLAS survey area at $z\sim 3$ (big stars); filled stars refer to far-IR detections with \textsl{Herschel} and empty symbols to stacked measurements (see Table 3 for more detailed information on photometric redshift). The thin solid black line is the average relationship obtained 
according to the AGN variability model \citep[][see text for details]{Hickox2014,Stanley2015,Stanley2017}.}\label{fig:coevplane}
\end{figure*}

We now turn to physically interpret our results in terms of an in-situ coevolution scenario between the host galaxy and the central SMBH, developed by \citet{Lapi2014,Lapi2017} \citep[see also][]{Mancuso2016,Mancuso2017}. To such purpose, we focus on the SFR vs. $L_{\rm AGN}$ plane, namely, a diagram where the SFR in the host galaxy is plotted against the bolometric AGN luminosity $L_{\rm AGN}$, see Fig.~\ref{fig:coevplane}. 

The dotted line represents the locus where the bolometric AGN luminosity $L_{\rm AGN}$ equals the bolometric luminosity $L_{\rm SFR} [{\rm erg~s}^{-1}]\sim 3\times 10^{43}\,$ SFR $[M_\odot~{\rm yr}^{-1}]$ associated to star formation in the host. In our scenario  \citep[see][for details]{Lapi2014,Mancuso2016} galaxies located to the right of such a line have accumulated appreciable stellar masses, host a growing SMBH that has attained considerable luminosities, and so are prone to be strongly affected by the resulting AGN energy/momentum feedback.
The dashed lines illustrate three typical evolutionary tracks for individual galaxies based on our coevolution scenario, which correspond to values of SFR $\sim 400$, $1200$, and $4000\, M_\odot$ yr$^{-1}$. Note that these tracks refer to the time-average behavior of the AGN luminosity over $100$ Myr on more, but do not exclude the presence of variability over much shorter timescales.

During the early stages of a massive galaxy's evolution, the SFR is approximately constant with the galactic age, while the AGN luminosity is smaller than that associated to star formation, but increases much more rapidly; in the SFR vs. $L_{\rm AGN}$ plane this generates an evolutionary track which is closely parallel to the $L_{\rm AGN}$ axis. After some $10^8$ yr, the nuclear power progressively increases to values overwhelming that from star formation in the host, which is affected and reduced on a rapid timescale, and soon after even the AGN luminosity decreases because fuel in the central region gets exhausted; in the SFR vs. $L_{\rm AGN}$ plane this generates a track receding toward the bottom left portion of the diagram.

The average statistical relationship between SFR and AGN luminosity in the SFR vs. $L_{\rm AGN}$ plane can be determined on the basis of these tracks for individual galaxies, by taking into account three additional ingredients: (i) the number density of galaxies with given value of the SFR, implying that galaxies with higher SFRs are rarer; (ii) the duty cycle, specifying that a galaxy spends different relative time intervals in the different portions of the track; (iii) the detection threshold for star formation in the host by current far-IR measurements, amounting to SFR$\sim 10^2\, M_\odot$ yr$^{-1}$. The thick solid lines shows the resulting average relationship and the shaded areas render its $1\sigma$ (light grey), $2\sigma$ (grey), and $3\sigma$ (dark grey) variance \citep[see][for details on the computation]{Mancuso2016}. The rise of the average SFR for increasing values of $L_{\rm AGN}$ just occurs because, statistically, to achieve a higher AGN luminosity the BH must reside in a more massive galaxy with higher initial SFR.

For comparison, we report in the coevolution plane the literature data from \citet[circles]{Stanley2015,Stanley2017} at $z\sim 2$, \citet[squares]{Netzer2016} at $z\sim 3$, \citet[pentagons]{Venemans2018} at $z\sim 6$, \citet[inverse triangles]{Fan2016} at $z\sim 3$, \citet[triangles]{Xu2015} at $z\sim 2$ and \citet[diamonds]{Page2012} at $z\sim 2$; the SFRs have been converted to a Chabrier IMF \citep[see Table 2 in][for conversion factors]{Bernardi2010}. Red symbols refer to X-ray selected AGNs, blue to optically-selected AGNs, and green to mid-IR selected AGNs; the SFR measurements are obtained from far-IR observations with \textsl{Herschel}, both from detections (filled symbols) and from stacking techniques (empty symbols); the sample by \citet{Venemans2018} is based on detections from ALMA millimetric data. For example, in the upper right portion of the diagram, the mid-IR selected objects by \citet{Fan2016} are hyper-luminous HotDOGs; these are violently starforming galaxies featuring a high-temperature dust component likely heated up by the powerful energy/momentum feedback from the central AGN.

According to our scenario outlined above, the position on the diagram of these data points can be understood as follows. Optical data tend to pick up objects close to the peak of AGN luminosity when dust has been partially removed and the SFR is being progressively affected; X-ray data can pick up objects before or after the AGN peak, hence with dusty SFR in the host still sustained or reduced, respectively. Mid-IR data with the current observational limits strike an intermediate course between the former two. On top of that, detections in the far-IR with \textit{Herschel} tend to probe galaxies when the SFR is still sustained to high values, while stacked measurements tend to probe the epochs after the SFR has been partly affected by the AGN feedback. Very interestingly, the ALMA millimetric data by \citet{Venemans2018} in powerful quasars at $z\sim 6$ start to detect the star formation in the host even after its suppression has already started/occurred.

Finally, results from this work are illustrated as magenta stars. In order of increasing symbol size, we show the outcome for the overall sample at $\langle z\rangle\sim 1.5 - 2$ (tiny stars), for \textsl{WISE} $W4$-detected galaxies at $\langle z\rangle\sim 1.5 - 2$ (small stars), for \textsl{WISE} $W4$-detected galaxies with $S_{W4}/S_{r}$ ratio exceeding $100$ at $\langle z\rangle\sim 2-3$ (medium-size stars), and for the unique HotDOG falling in the \textsl{Herschel}-ATLAS survey area (big stars) with $S_{W4}/S_{r}$ ratio exceeding $1000$, which is located at $z\sim 3$; filled stars refer to far-IR detections with \textsl{Herschel} and empty symbols to the corresponding stacked measurements.

We can draw the following picture. First, our results well follow the average SFR vs. $L_{\rm AGN}$ relationship, within its $1\sigma$ dispersion; actually we can identify two parallel sequences, one for detected sources which lies above the mean relationship, and one for stacked sources which lies below. It is remarkable that, despite our sample is primarily pre-selected in the optical band from the SDSS survey, our additional selection involving mid-IR bands (i.e., requiring detection in WISE W4 band, or $S_{W4}/S_{r}\ga 100$ or prominent emission in $S_{W4}/S_{r}\ga 1000$ as in the HotDOG) constitute an effective way to move along the average relationship toward increasing AGN luminosities.

Second, in terms of evolutionary track for individual galaxies, the results of this work are located just before (average for far-IR/(sub-)mm detected sources) or just after (average for undetected sources) the epoch when the star formation starts to be strongly affected by the AGN feedback. Note that, given the rarity of such hyper-luminous objects and the relatively small \textsl{Herschel}-ATLAS survey area, it is problematic to pinpoint the stacked measurements corresponding to the HotDOG. \citet{Stacey2018}, albeit with noticeable uncertainties, has tentatively detected with \textsl{Herschel} suppressed star formation in powerful quasar hosts at $z\sim 3$ thanks to gravitational lensing by foreground objects. At much higher redshift $z\sim 6$, \citet{Venemans2018} has found similar suppressed star formation in powerful quasar hosts thanks to millimeter ALMA observations. In the future it would be extremely interesting to complement these data with information from X-ray surveys, that could allow to probe the individual evolutionary tracks both in the early stages of obscured BH accretion and in the late stage after the onset of feedback.

Finally, it is worth reporting that another popular interpretation of the SFR vs. $L_{AGN}$ diagram calls into play AGN variability \citep[see][]{Hickox2014,Aird2013,Stanley2015,Stanley2017}, as inspired  by merger-driven models of galaxy formation and numerical simulations  \citep[e.g.][]{Novak2011,Veale2014,Volonteri2015,McAlpine2017}. Such a scenario envisages that, when averaged over timescales longer than 100 Myr,  the AGN luminosity $\langle L_{AGN}\rangle$ is correlated to the SFR so that to keep a constant ratio $1/3000$ of the BH accretion rate to the SFR, as observed in the local universe \citep[e.g.][]{Chen2013}). However, the luminosities $L_{AGN}$ of individual AGNs are allowed to vary on shorter timescales according to an assumed AGN luminosity distribution ${\rm d}p/{\rm d}\log L_{AGN}\propto L_{AGN}^{−0.2}\, \exp{(-L_{AGN}/100\,\langle L_{AGN}\rangle)}$ which features a broken power-law form at the faint end and an exponential cutoff. Coupling such a distribution with the statistics of galaxies of given SFR and redshift yields an average relationship in the SFR vs. $L_{AGN}$ plane. We perform the computation with the variability model by using the same SFR galaxy statistics and detection threshold employed above for our scenario, and show in Fig. \ref{fig:coevplane} the outcome as a thin solid line. With the present data
it is challenging to differentiate between the in-situ coevolution and the variability scenarios. 
A marked difference can be recognized in the statistics and the interpretation of the objects in the bottom right region of the
SFR vs. $L_{AGN}$ plane, i.e., at high AGN luminosity $L_{AGN}\gtrsim$ some $10^{47}$ erg s$^{-1}$ and 
relatively low SFRs $\la$ a few $10^2\, M_\odot$ yr$^{-1}$. 
In the in-situ coevolution picture, these are galaxies
close to the end of their evolution, when strong AGN activity powered by a BH with mass $\gtrsim 10^9\, M_\odot$ 
is removing dust and suppressing the SFR with respect to the 
high values $\ga 10^3\, M_\odot$ yr$^{-1}$ applying to the early dust-enshrouded phase; 
as such these systems should feature large stellar masses $M_\star\gtrsim 10^{11}\, M_\odot$ already accumulated 
in their central regions $\la$ a few Kpcs.
Contrariwise, in the AGN variability framework, these galaxies 
are selected by chance at AGN luminosities substantially higher than the average because of short-time variability;
the adopted AGN luminosity distribution will limit their number density and will imply independence on the stellar mass. In the near future, accurate estimates of the number density of the galaxies populating the bottom right region of the 
SFR vs. $L_{AGN}$  plane, and precise measurements of the 
stellar masses in these systems will contribute to distinguish between the two competing scenarios; these
observations will likely become feasible, even at substantial redshifts, with the advent  
of the James Webb Space Telescope.

\subsection{Future outlook}
We now move on to discuss natural extensions of the analysis presented in this paper. \\
A promising avenue that has recently emerged to investigate the AGN activity -- and in particular the feedback energetics -- is its thermal Sunyaev-Zel'dovich (tSZ) signature \citep{SZ1970,SZ1972} at millimeter wavelengths. The energy released by the AGN to the surrounding diffuse baryons affects their temperature and distribution. As a consequence, the heated electron gas can in principle Compton-upscatter the cosmic microwave background (CMB) photons, leaving an imprint in the CMB blackbody spectrum by tilting it to higher frequencies  \citep{Voit1994,Natarajan1999,Lapi2003,Chatterjee2007,Scannapieco2008}. However, the AGN feedback-induced tSZ is expected to be too feeble to be detected at the single source level, therefore one must resort to statistical techniques such as cross-correlation and stacking. The first tentative detection of tSZ due to AGN feedback effect was reported by  \citet{Chatterjee2009}, while subsequent works by \citet{Ruan2015,Crichton2016,Verdier2016,Spacek2016,Spacek2017,soergel:2017} found evidence of this effect at different significance levels. However, there is no general consensus as the inferred thermal energies differ substantially among the various studies. One of the main obstacle in this analysis is represented by the strong dust signal which dominates the SED at $\nu \gtrsim 100$ GHz \citep{Crichton2016,Verdier2016,soergel:2017} and creates a degeneracy with the tSZ signal. An additional complication involves distinguishing between the AGN-induced signal and that of the virialized gas in the host dark matter halo, see e.g. \citet{Chowdhury2017}.\\
Following \citet{soergel:2017}, in this paper we have moved the first steps in addressing this aspect by stacking maps at \textit{Planck} wavelengths (above and below the SZ null frequency at $\simeq 217$ GHz). One of the outcome of this work is also to inform us about the best approach and most suited AGN samples to investigate the AGN feedback-induced tSZ imprint which is ultimately limited by the dust emission. A possible way forward to look for this SZ signature would be to use high-resolution and high-sensitivity ground-based telescopes with multi-frequency coverage.

\section{Conclusions}
\label{sec:conclusions}

In this paper we have provided a multi-wavelength picture of SDSS optically-selected QSOs in the redshift range $1 \le z \le 5$ from the near-UV to the sub-mm bands. To this end, we have combined diverse photometric information from SDSS, UKIDSS, and WISE surveys, as well as performing a stacking analysis on \textit{Herschel}, \textit{AKARI}, and \textit{Planck} maps. 

The shape of the reconstructed mean SED is similar to other templates in the UV-optical part of the spectrum, while it features a little excess at 1-2 $\mu$m (likely due to old stars and heated dust), a prominent bump around 4-6 $\mu$m followed by a decrease out to $\sim 20 \mu$m and a subsequent far-IR bump comparable to what measured by other studies. By decomposing the SED into an AGN, starburst, and torus components, we have estimated the AGN luminosity and SFR, finding typical $L_{AGN} \sim 10^{46} - 10^{47}$ erg/s and SFR $\sim 50 - 1000\, M_{\odot}/$yr. 

When interpreted in the context of the in-situ co-evolution scenario for galaxies and hosted SMBHs presented by \citet{Lapi2014}, our results suggest that the mid-IR based selection criteria developed in this work (detection in the WISE $W4$ band and $S_{W4}/S_{r}>100$) have been proven effective in moving along the SFR vs. $L_{\rm AGN}$ diagram towards higher AGN luminosities; on the other hand, the far-IR/sub-mm detection criterion, if (not) enforced, allows us to probe the epoch just (after) before the onset of feedback from the central AGN.

Natural extensions of the current work involve complementing these datasets with X-ray observations, in order to shed light on the early stages of the BH accretion and the evolution after the onset of AGN feedback, as well as with mm and radio information to investigate the alleged AGN feedback-induced Sunyaev-Zel'dovich emission and constrain the non-gravitational heating of the Intergalactic Medium.

\acknowledgments{We thank the referee for constructive comments that helped in improving the paper. We are indebted to Luigi Danese for his support during the early stages of this work. We acknowledge very fruitful discussions with S. Mateos and F.J. Carrera. We also thank Simone Aiola, Srinivasan Ragunathan, Christian L. Reichardt and Reinier Janssen for providing comments on an earlier version of the manuscript. FB acknowledges support from an Australian Research Council Future Fellowship (FT150100074). GF acknowledges the support of the CNES postdoctoral program. JGN acknowledges financial support from the I+D 2015 project AYA2015- 65887-P (MINECO/FEDER) and from the Spanish MINECO for a "Ramon y Cajal" fellowship (RYC-2013-13256). AL acknowledges the RADIOFOREGROUNDS grant (COMPET-05-2015, agreement no. 687312) of the European Union Horizon 2020 research and innovation program and the MIUR grant "Finanziamento annuale individuale attivit\'a base di ricerca". RG acknowledges support from the agreements ASI-INAF n.2017-14-H.O and ASI-INAF I/037/12/0. \\
\\
Funding for SDSS-III has been provided by the Alfred P. Sloan Foundation, the Participating Institutions, the National Science Foundation, and the U.S. Department of Energy Office of Science. The SDSS-III web site is \url{http://www.sdss3.org/}. SDSS-III is managed by the Astrophysical Research Consortium for the Participating Institutions of the SDSS-III Collaboration including the University of Arizona, the Brazilian Participation Group, Brookhaven National Laboratory, Carnegie Mellon University, University of Florida, the French Participation Group, the German Participation Group, Harvard University, the Instituto de Astrofisica de Canarias, the Michigan State/Notre Dame/JINA Participation Group, Johns Hopkins University, Lawrence Berkeley National Laboratory, Max \textit{Planck} Institute for Astrophysics, Max \textit{Planck} Institute for Extraterrestrial Physics, New Mexico State University, New York University, Ohio State University, Pennsylvania State University, University of Portsmouth, Princeton University, the Spanish Participation Group, University of Tokyo, University of Utah, Vanderbilt University, University of Virginia, University of Washington, and Yale University.
}

\software{Astropy \citep{astropy:2013,astropy:2018}, IPython \citep{ipython}, matplotlib \citep{Hunter:2007}, scipy \citep{scipy}, HEALPix \citep{healpix} }

\bibliographystyle{yahapj}
\bibliography{references}

\appendix
\section{Testing the far-IR sub-sample representativity}
\label{sec:HATLAS_subsample_check}
\begin{figure*}[htbp]
\includegraphics[width=\textwidth]{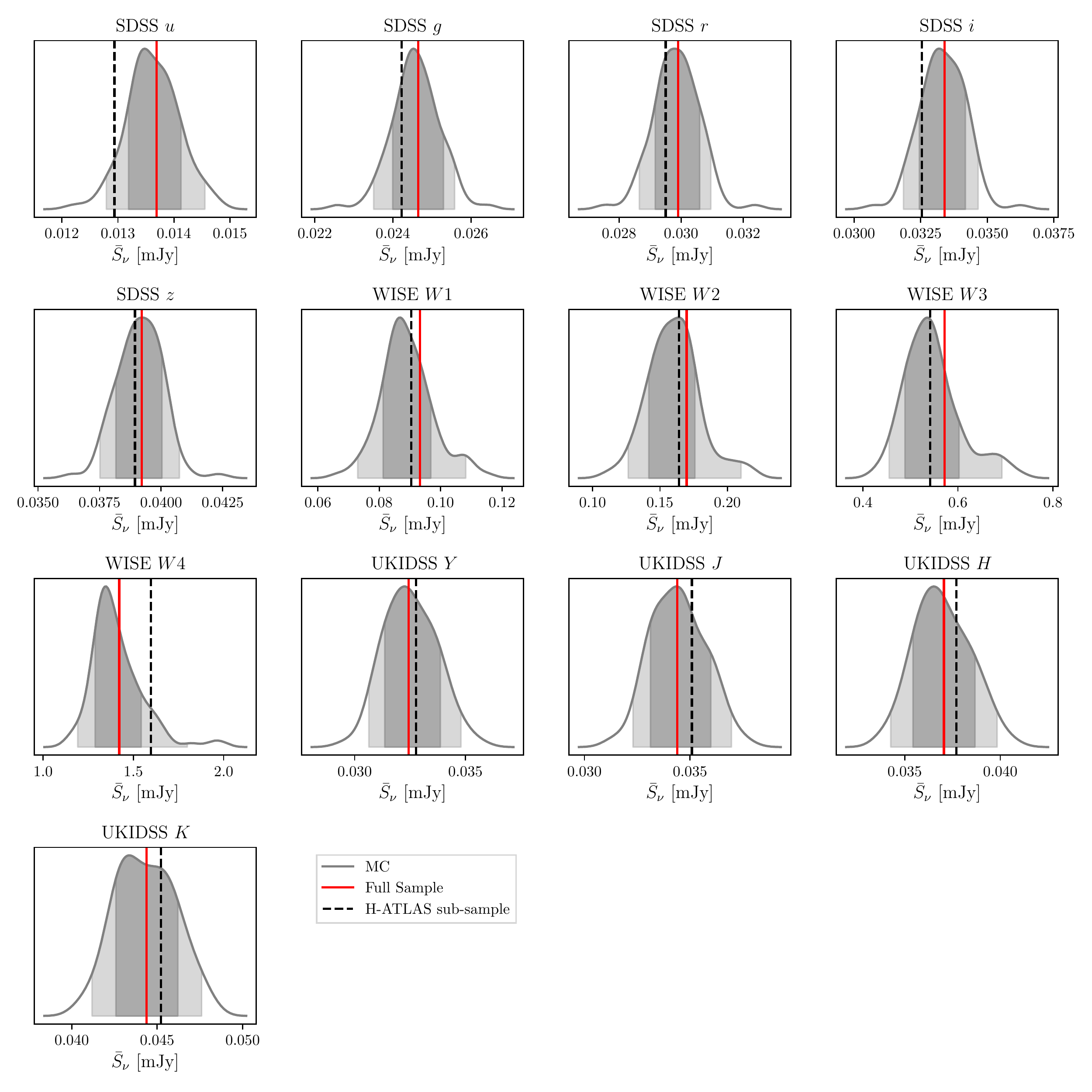}
    \caption{Assessment of the H-ATLAS sub-sample representativity. In each panel, the grey line shows a kernel density estimate of the distribution of the mean fluxes in a certain photometric band estimated from a set of 100 random sub-samples. The darker and lighter shaded areas indicate the $\pm 1\sigma$ and $\pm 2 \sigma$ of the average flux distribution, evaluated as the (13.6, 86.4) and (2.1, 97.9) percentiles of the distribution respectively. The solid red and dashed black lines denote the average flux measured for whole QSO sample and the H-ATLAS sub-sample respectively.}
    \label{fig:subsample_check}
\end{figure*}

In this appendix, we assess to what extent the fluxes of sources laying in the H-ATLAS footprint (i.e. the H-ATLAS sub-sample) are representative of the entire QSO sample. Adopting representative sub-samples is in fact a crucial point when studying the demographics of a given sources population. \\*
As discussed in Sec.~\ref{sec:samples}, when no sub-mm detection criterion is enforced, the average photometry for the optical and mid-IR part of the SED is calculated over the whole QSO sample, while the far-IR/sub-mm portion of the spectrum is estimated using only sources included in the H-ATLAS footprint. The main reason for this strategy is simply to increase the sample size statistics, an issue that mostly affects the low resolution and relatively high detection limits experiments such as \textit{AKARI} and \textit{Planck}. However, one might wonder whether the fluxes measured for the H-ATLAS sub-sample are representative of the full sample and therefore provide a reliable estimation of the average flux. In order to test this assumption, we proceed in the following way. From the main QSO sample,\footnote{Note that, for the purposes of this check, we consider all the sources within the AllQSO sample in a single redshift bin $1 \le z \le 5$.} we randomly draw $N_{\rm MC} = 100$ sub-samples containing a number of sources comparable to that of the H-ATLAS one (about 7000 sources, or the 3\% of the whole QSO sample with SDSS fluxes). Then, from each of these 100 random sub-sample, we estimate the average flux $\bar{S}_{\nu}$ distribution in each photometric band and compare it to the average of fluxes measured firstly in the H-ATLAS sub-sample and secondly over the whole SDSS footprint. The results of this test are shown in Fig.~\ref{fig:subsample_check}. In each panel, the grey line shows the distribution of the mean fluxes across the random sub-samples at a given photometric band, while the solid red and dashed black lines denote the average flux measured for the whole QSO sample and the H-ATLAS sub-sample respectively. The darker and lighter shaded regions indicate the confidence levels equivalent to the $\pm 1\sigma$ and $\pm 2 \sigma$ of a Gaussian distribution for the average flux distribution.\footnote{Evaluated as the (13.6, 86.4) and (2.1,97.9) percentiles of the distribution respectively.} The average fluxes measured within the H-ATLAS fields do not appear to be outliers of the distribution, as they fall within the $\pm 1\sigma$ band. The mean flux in 2 out of the 13 photometric band are within the $\pm 2 \sigma$, but approximately $1/3$ of the measurements are expected to be scattered more than 1$\sigma$. Hence, we conclude that we can safely consider the far-IR/sub-mm average fluxes measured in the H-ATLAS fields to be representative of the whole QSO sample.

\section{SED templates}
In this appendix section we report the best-fit SEDs for main QSO sample in the three redshift bins  between $1 \le z \le 5$. The tabulated values as function of the rest-frame wavelength are presented in Tab.~\ref{tab:sed_fit_allqso}.

\begin{table}[ht]
\centering
\caption{Tabulated values of the overall average SED for the baseline optical-selected QSO sample as function of redshift and in terms of the rest-frame wavelength.}
\label{tab:sed_fit_allqso}
\begin{tabular}{c|c|c|c}
                                & \multicolumn{3}{c|}{$S_{\nu}$ {[}mJy{]}}                              \\ \cline{2-4} 
$\log_{10}\lambda \, [\micron]$ & $1<z<2.15$            & $2.15 < z < 2.5$      & $2.5 < z < 5$         \\ \hline
$-1.0$                          & $1.67 \times 10^{-2}$ & $1.33 \times 10^{-2}$ & $1.49 \times 10^{-2}$ \\
$-0.9$                          & $2.25 \times 10^{-2}$ & $1.79 \times 10^{-2}$ & $2.00 \times 10^{-2}$ \\
$-0.8$                          & $2.80 \times 10^{-2}$ & $2.23 \times 10^{-2}$ & $2.50 \times 10^{-2}$ \\
$-0.7$                          & $3.38 \times 10^{-2}$ & $2.69 \times 10^{-2}$ & $3.01 \times 10^{-2}$ \\
$-0.6$                          & $4.12 \times 10^{-2}$ & $3.28 \times 10^{-2}$ & $3.66 \times 10^{-2}$ \\
$-0.5$                          & $4.40 \times 10^{-2}$ & $3.50 \times 10^{-2}$ & $3.92 \times 10^{-2}$ \\
$-0.4$                          & $4.05 \times 10^{-2}$ & $3.21 \times 10^{-2}$ & $3.60 \times 10^{-2}$ \\
$-0.3$                          & $3.78 \times 10^{-2}$ & $3.03 \times 10^{-2}$ & $3.40 \times 10^{-2}$ \\
$-0.2$                          & $4.23 \times 10^{-2}$ & $3.49 \times 10^{-2}$ & $3.96 \times 10^{-2}$ \\
$-0.1$                          & $5.40 \times 10^{-2}$ & $4.56 \times 10^{-2}$ & $5.29 \times 10^{-2}$ \\
$0.0$                           & $7.85 \times 10^{-2}$ & $6.29 \times 10^{-2}$ & $7.45 \times 10^{-2}$ \\
$0.1$                           & $1.18 \times 10^{-1}$ & $8.35 \times 10^{-2}$ & $9.88 \times 10^{-2}$ \\
$0.2$                           & $1.76 \times 10^{-1}$ & $1.09 \times 10^{-1}$ & $1.21 \times 10^{-1}$ \\
$0.3$                           & $2.64 \times 10^{-1}$ & $1.68 \times 10^{-1}$ & $1.61 \times 10^{-1}$ \\
$0.4$                           & $3.73 \times 10^{-1}$ & $2.83 \times 10^{-1}$ & $2.68 \times 10^{-1}$ \\
$0.5$                           & $4.70 \times 10^{-1}$ & $4.27 \times 10^{-1}$ & $4.82 \times 10^{-1}$ \\
$0.6$                           & $6.02 \times 10^{-1}$ & $5.61 \times 10^{-1}$ & $7.97 \times 10^{-1}$ \\
$0.7$                           & $9.05 \times 10^{-1}$ & $7.45 \times 10^{-1}$ & $1.13 \times 10^{0}$  \\
$0.8$                           & $1.35 \times 10^{0}$  & $1.00 \times 10^{0}$  & $1.30 \times 10^{0}$  \\
$0.9$                           & $1.69 \times 10^{0}$  & $1.23 \times 10^{0}$  & $1.23 \times 10^{0}$  \\
$1.0$                           & $1.70 \times 10^{0}$  & $1.27 \times 10^{0}$  & $9.86 \times 10^{-1}$ \\
$1.1$                           & $1.42 \times 10^{0}$  & $1.10 \times 10^{0}$  & $6.90 \times 10^{-1}$ \\
$1.2$                           & $1.03 \times 10^{0}$  & $8.18 \times 10^{-1}$ & $4.37 \times 10^{-1}$ \\
$1.3$                           & $6.97 \times 10^{-1}$ & $5.60 \times 10^{-1}$ & $2.77 \times 10^{-1}$ \\
$1.4$                           & $6.40 \times 10^{-1}$ & $4.85 \times 10^{-1}$ & $3.05 \times 10^{-1}$ \\
$1.5$                           & $1.19 \times 10^{0}$  & $8.06 \times 10^{-1}$ & $7.18 \times 10^{-1}$ \\
$1.6$                           & $2.41 \times 10^{0}$  & $1.57 \times 10^{0}$  & $1.56 \times 10^{0}$  \\
$1.7$                           & $3.83 \times 10^{0}$  & $2.57 \times 10^{0}$  & $2.52 \times 10^{0}$  \\
$1.8$                           & $5.01 \times 10^{0}$  & $3.22 \times 10^{0}$  & $3.31 \times 10^{0}$  \\
$1.9$                           & $5.84 \times 10^{0}$  & $3.75 \times 10^{0}$  & $3.87 \times 10^{0}$  \\
$2.0$                           & $6.11 \times 10^{0}$  & $3.92 \times 10^{0}$  & $4.04 \times 10^{0}$  \\
$2.1$                           & $5.53 \times 10^{0}$  & $3.56 \times 10^{0}$  & $3.67 \times 10^{0}$  \\
$2.2$                           & $4.29 \times 10^{0}$  & $2.76 \times 10^{0}$  & $2.84 \times 10^{0}$  \\
$2.3$                           & $2.88 \times 10^{0}$  & $1.85 \times 10^{0}$  & $1.91 \times 10^{0}$  \\
$2.4$                           & $1.71 \times 10^{0}$  & $1.10 \times 10^{0}$  & $1.14 \times 10^{0}$ 
\end{tabular}
\end{table}

\section{SED fitting methodology test\label{sec:sed_test}}
When dealing with stacked sources (in our case their spectroscopic redshift is known but not their individual flux densities) there is always an assumption to make in order to extract useful information from the analyzed data. The methodology used for the SED fitting in this work assumes that the different adopted SED templates are representative of the whole population inside the redshift bin. Therefore, they are simply used to estimate the main redshift of the sample and the flux density normalization to eventually derive the intrinsic luminosities (see section \ref{sec:sed_model} for more details).

However, we caveat that high redshift objects ($z > 1$) can suffer strong $K-$corrections. One may wonder whether our results are stable when taking into account the individual redshifts of the sources, so avoiding to mix information from different wavelengths. For example, this is the approach followed by \citet{Crichton2016} in the IR band. These authors consider a grey-body emission with emissivity index $\beta$ and dust temperature as fitting parameters; on the other hand, they are forced to assume the same luminosity for all the object inside the redshift bin. Then they estimate the stacked redshifted SEDs following the same procedure applied to the data and use it to get the best SED normalization (in terms of bolometric luminosity).

In order to check the robustness of our results, we perform a simple test comparing both approaches. As model SED we use the averaged SED estimated for the All QSO sample in the first redshift bin (the most critical one since the redshift distribution there is rather flat). We use the full SED in order to show the main differences between the methodologies, only visible in the most prominent SED features.

\begin{figure*}[htbp]
\includegraphics[width=\textwidth]{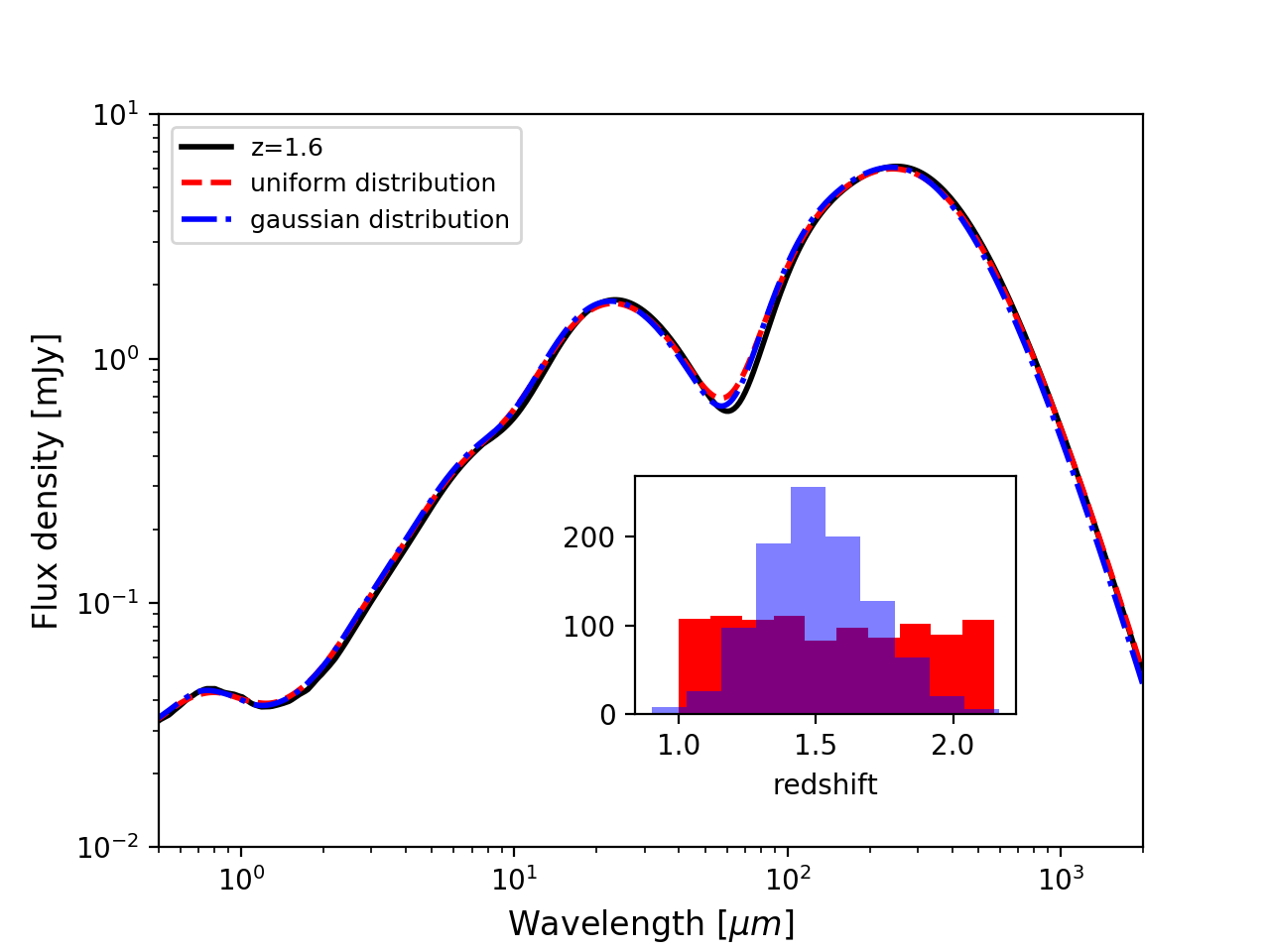}
    \caption{Results of the simple test using two different methodologies to estimate the average SED (see text for more details): main sample of QSOs in the first redshift bin ($1.0 < z < 2.15$) derived at the mean redshift ($z=1.6$, \textit{black solid line}); rest-frame averaged SED considering an uniform (\textit{dashed red line}) and a gaussian (\textit{dot-dashed blue line}) redshift distributions inside the redshift bin.}
    \label{fig:sed_test}
\end{figure*}

We consider two simulated populations of 1000 QSOs with two different redshift distributions (see inset in Fig.~\ref{fig:sed_test}): an uniform distribution (\textit{red histogram}) and a gaussian distribution (\textit{blue histogram}). Taking into account the different redshifts for each QSO, we then estimate the main redshifted SED for both redshift distributions. We finally compare them with the initial SED at the mean redshift $z=1.6$.

As can be observed in Fig.~\ref{fig:sed_test}, the three estimated SEDs are almost identical. The main effect of considering a redshift distribution is a negligible smoothing, mostly observed in the valley feature around $60\, \mu m$ at $z=1.6$. Therefore, we can safely conclude that our results and conclusions are robust and independent of the adopted SED fitting methodology.

\end{document}